\documentclass[journal]{IEEEtran}
\usepackage{cite}

\ifCLASSINFOpdf
  \usepackage[pdftex]{graphicx}
\else
  \usepackage[dvips]{graphicx}
\fi

\usepackage{amsmath}

\usepackage{algorithmicx}
\usepackage{algorithm}
\usepackage{algpseudocode}

\usepackage{array}

\usepackage{fixltx2e}

\usepackage{url}

\usepackage{bbm, bm}
\usepackage[table]{xcolor}
\usepackage{hyperref}
\usepackage{tabularx}
\usepackage{makecell}
\usepackage{booktabs}
\usepackage{amssymb}
\usepackage{mathtools}
\usepackage{tcolorbox}

\usepackage{wrapfig}
\usepackage{subfigure}
\DeclareMathOperator*{\sign}{sign}
\DeclareMathOperator*{\BLER}{BLER}
\DeclareMathOperator*{\BER}{BER}

\DeclarePairedDelimiter\floor{\lfloor}{\rfloor}
\newcommand{\Mod}[1]{\ (\mathrm{mod}\ #1)}

\begin{document}
\title{TransCoder: A Neural-Enhancement Framework\\ for Channel Codes\\}

\author{Anastasiia~Kurmukova,~\IEEEmembership{Student~Member,~IEEE,}
        Selim~F.~Yilmaz,
        Emre~Ozfatura,~\IEEEmembership{Member,~IEEE,}
        %~\IEEEmembership{Fellow,~OSA,}
        and~Deniz~G{\"u}nd{\"u}z,~\IEEEmembership{Fellow,~IEEE,}% <-this % stops a space
\thanks{A. Kurmukova is with the Department
of Electrical and Electronic Engineering, Imperial College London, London SW7 2AZ, U.K. \textit{(Corresponding author.)} E-mail: \href{mailto:a.kurmukova22@imperial.ac.uk}{a.kurmukova22@imperial.ac.uk}.}% <-this % stops a space
\thanks{S. F. Yilmaz was with the Department
of Electrical and Electronic Engineering, Imperial College London, London SW7 2AZ, U.K., when the work was done. E-mail: \href{mailto:s.yilmaz21@imperial.ac.uk}{s.yilmaz21@imperial.ac.uk}.}%
\thanks{Emre Ozfatura is with the Department
of  Electronics Engineering, Sabanci University, Istanbul, Turkey. E-mail: \href{mailto:emre.ozfatura@sabanciuniv.edu}{emre.ozfatura@sabanciuniv.edu}.}
\thanks{D. G{\"u}nd{\"u}z is with the Department
of Electrical and Electronic Engineering, Imperial College London, London SW7 2AZ, U.K. E-mail: \href{mailto:d.gunduz@imperial.ac.uk}{d.gunduz@imperial.ac.uk}.}%
% \thanks{Manuscript received November 27, 2025.}
}

\maketitle

\begin{abstract}
Reliable communication over noisy channels requires the design of specialized error-correcting codes (ECCs) tailored to specific system requirements. Recently, neural network-based decoders have emerged as promising tools for enhancing ECC reliability, yet their high computational complexity prevents their potential practical deployment. In this paper, we take a different approach and design a neural transmission scheme that employs the transformer architecture in order to improve the reliability of existing ECCs. We call this approach TransCoder, alluding both to its function and architecture. TransCoder operates as a code-adaptive neural module aimed at performance enhancement that can be implemented flexibly at either the transmitter, receiver, or both. The framework employs an iterative decoding procedure, where both noisy information from the channel and updates from the conventional ECC decoder are processed by a neural decoder block, utilizing a block attention mechanism for efficiency. Through extensive simulations with various conventional codes (LDPC, BCH, Polar, and Turbo) and across a wide range of channel conditions, we demonstrate that TransCoder significantly improves block error rate (BLER) performance while maintaining computational complexity comparable to traditional decoders. Notably, our approach is particularly effective for longer codes (block length $>64$) and at lower code rates, scenarios in which existing neural decoders often struggle (despite their formidable computational complexity). The results establish TransCoder as a promising practical solution for reliable communication among resource-constrained wireless devices.
\end{abstract}

\begin{IEEEkeywords}
Error correction codes, Block-attention mechanism, Transformer, Wireless communication, Neural decoder.
\end{IEEEkeywords}

\IEEEpeerreviewmaketitle

\section{Introduction}

\IEEEPARstart{R}{eliable} communication over noisy channels is essential in modern digital systems, and error-correcting codes (ECCs) play a critical role in ensuring data integrity. Classical linear codes, such as polar codes \cite{polar} and low-density parity-check (LDPC) codes \cite{LDPC}, and linear codes with memory, such as Turbo codes \cite{turbo}, which are based on convolutional codes \cite{elias}, have been shown to achieve near-capacity performance \cite{LDPC_shannon} on additive white Gaussian noise (AWGN) channels for long block lengths. Although these codes have been successfully adopted in modern communication network protocol \cite{3gpp}, there are various open challenges that continue to drive research in this area, including suboptimal performance in short block length regimes \cite{bounds_polyanskiy}, limited adaptability to varying channel conditions \cite{rayleigh}, and the rapid growth of maximum-likelihood decoding complexity with the block length.

In particular, decoding complexity is a critical metric in evaluating ECCs for deployment in practical wireless systems. Since optimal maximum likelihood decoding for linear codes is NP-hard, practical implementations rely on efficient iterative decoding algorithms tailored to specific code families. These include Tanner graph based belief propagation (BP) for LDPC codes \cite{LDPC}, factor graph based BP \cite{bp_polar1, bp_polar2}, successive cancellation (SC) \cite{polar} and successive cancellation list (SCL) \cite{scl_polar} decoding for polar codes, and the Bahl-Cocke-Jelinek-Raviv (BCJR) decoding algorithm for Turbo codes \cite{turbo}. Demand for low-complexity near-optimal decoding solutions, coupled with advances in neural network architectures, has led to many recent studies on neural-assisted error correction, with most of the efforts focusing on improving the decoding performance of existing codes. Neural BP reduces the bit error rate (BER) by introducing trainable weights into conventional BP decoding on Tanner graphs \cite{neural_bp1, neural_bp2}, the polar factor graph \cite{neural_minsum, neural_bp_polar}, or the Turbo decoder \cite{tinyturbo}. Further replacement of conventional algorithms with trainable architectures, while preserving the decoding structure, resulted in models such as neural successive cancellation \cite{neural_polar_haim} for polar codes and neural BCJR \cite{neuralbcjr} for convolutional codes.

\begin{figure*}[t!]
    \centering    \includegraphics[width=0.7\linewidth]{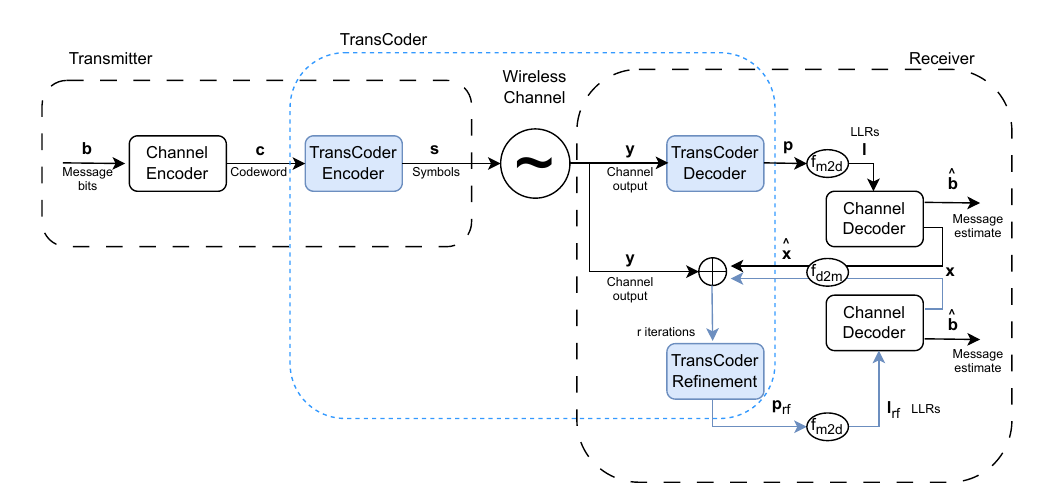}
    \caption{Overview of the TransCoder-based communication paradigm.}
    \label{fig:transcoder}
\end{figure*}

There are also fully neural decoding approaches \cite{on_deeplearning_based_cc} that are not based on any existing decoding algorithm but are limited to extremely short codes. The idea of syndrome-based neural decoding \cite{syndrom_based} for linear codes and the attention mechanism \cite{AttentionIsAllYouNeed} resulted in the creation of the error correction code transformer (ECCT) \cite{ECCT}, which embeds parity-check constraints into the self-attention mechanisms. This decoder surpasses the conventional BP decoder in terms of BER, and the model was further simplified through quantization and better masking in \cite{aecct}. Transformer-based decoding was further improved in \cite{crossmpt} by introducing cross-message passing. However, these fully transformer-based approaches result in huge increases in decoding complexity and drastic memory consumption compared to conventional decoders, which significantly limits their potential adoption in practical systems. Moreover, the improvements in BER often do not translate into improvements in the block error rate (BLER). Finally, it has been shown in \cite{Yuncheng:arXiv:24} that even the BER performance of these codes falls short of state-of-the-art (SoTA) decoding algorithms, casting serious doubt on the practical utility of these high-complexity ML-based decoding approaches.

Another approach is to modify the encoding function or the transmitted codeword of a fixed code while retaining the decoder. A good example of this approach is the method introduced in \cite{kurmukova2024friendly}, where the authors introduce minor perturbations to the modulated codeword to optimize its placement within the decision regions of the prescribed decoder, thereby improving its decoding accuracy without any modifications to the decoding algorithm. The standard modulation constellations can be improved without considering the channel code structure via signal shaping \cite{prob_geom_shaping} or through trainable signal shaping \cite{Jones:ECOC:19, Gumus:OFC:20, Aref:OFC:22}. A fully neural encoder and decoder without any pre-defined code structure \cite{intro_to_ml_cc} showed good performance only for extremely short code lengths. In \cite{KOcodes}, the authors introduced a neural architecture for both the encoder and decoder, preserving the recursive encoding/decoding structure of Reed-Muller codes, which demonstrates great error correction capability. However, such an approach requires considerable changes to the encoding/decoding procedure and is limited to one class of codes.

In this paper, we propose TransCoder, a novel approach that preserves the structure of existing ECCs while transforming the transmission scheme through neural processing at both ends of the communication link. Unlike previous neural approaches that modify the encoder or decoder directly, TransCoder operates as additional adaptive neural encoding/decoding modules (illustrated in Fig.~\ref{fig:transcoder}) that enhance the robustness of the transmitted codewords in standard BPSK/QPSK constellations. The complete TransCoder pipeline comprises three key components: a neural encoder module that adaptively adjusts codewords before noisy transmission, a primary decoder for initial channel output processing, and a refinement decoder that iteratively improves the noisy channel output. While the best performance is achieved with the full pipeline, an important advantage of our approach is its modular nature: even a partial implementation at either the transmitter or receiver can still yield substantial improvements.

\begin{figure}[h]
    \centering
    \includegraphics[width=.8\columnwidth]{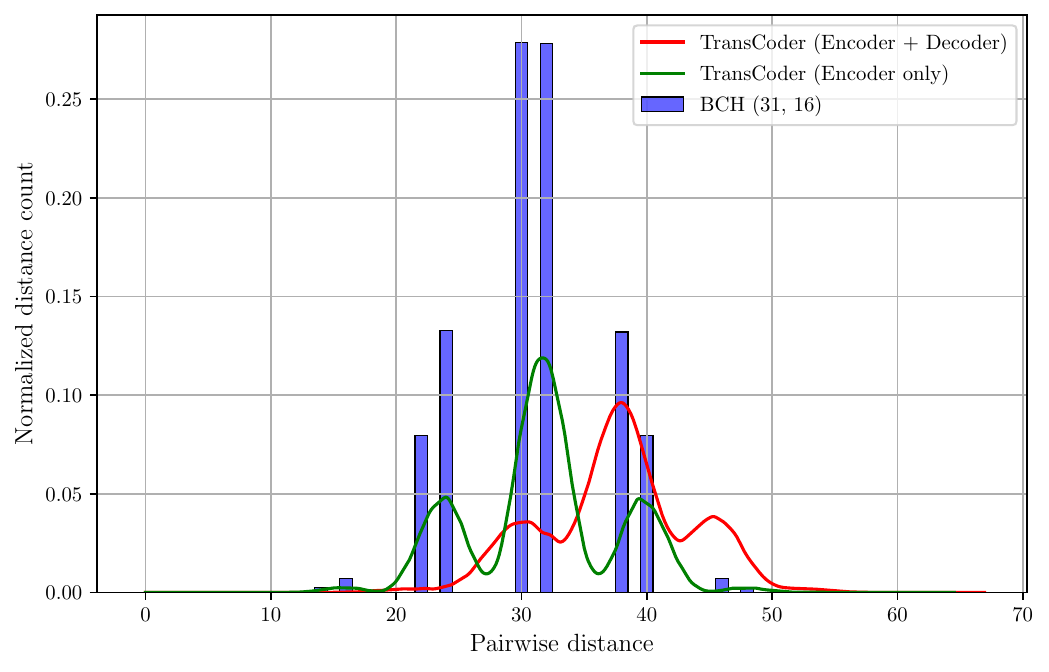}
    \caption{Normalized pairwise distances distribution between modulated codewords.}
    \label{fig:pairwise_distances}
\end{figure}

The effectiveness of TransCoder can be understood through its impact on the distribution of pairwise codeword distances. In Fig.~\ref{fig:pairwise_distances}, we plot the histogram of normalized pairwise distances between modulated codewords with and without TransCoder for the relatively short code BCH $(31,16)$ with just $2^{16}$ possible codewords. Our neural framework restructures the code distribution to enhance codeword distinguishability while preserving the underlying code structure. Even when employed only at the encoder, TransCoder achieves a more Gaussian-like distance distribution, similar to the theoretically optimal Gaussian codebook. Employing TransCoder at both ends results in a similar Gaussian-like distance distribution with an increased average pairwise distance.

Our comprehensive experimental evaluations employing various code families (LDPC, BCH, Polar, and Turbo) in diverse channel conditions demonstrate that TransCoder significantly outperforms both standard and neural-based decoders in terms of block error rate (BLER) performance. The gains of the proposed scheme are most pronounced for longer codes (with lengths greater than $64$ and up to $512$ that were tested) and lower code rates, while existing transformer-based approaches \cite{aecct, crossmpt} often show mediocre performance in these settings. For instance, with LDPC $(384,192)$ code, TransCoder achieves a $1$dB gain at a BLER of $10^{-5}$ compared to both conventional BP decoding with 20 iterations and CrossMPT \cite{crossmpt}. Having the conventional decoder incorporated inside the pipeline allows us to keep the transformer-based module small; thus, the TransCoder computational complexity exceeds that of conventional approaches like BP decoding by just one order of magnitude, while being two orders of magnitude less than the SoTA CrossMPT decoder.

These results represent a major milestone in the practicality of transformer-based error correction solutions for communication systems with limited computational and memory resources. By bridging the divide between SoTA neural decoding methods and the efficiency of traditional algorithms like BP, this work offers a compelling path toward real-world deployment in resource-constrained environments. Our major contributions can be summarized as follows.
\begin{itemize}

\item \textit{Novel Framework:} We introduce a novel end-to-end communication framework enabled through the TransCoder block, which effectively reshapes codeword distance distribution while preserving underlying code structures. In other words, TransCoder unifies traditional hand-crafted encoding/ decoding algorithms with neural network-assisted solutions, offering a robust and efficient error-correction framework.

\item \textit{Practical Efficiency:} TransCoder achieves superior error correction performance while maintaining computational complexity comparable to BP decoding, with flexible deployment options at the transmitter and/or the receiver.

\item \textit{Versatility:} The proposed TransCoder mechanism can be utilized in conjunction with all major code families (LDPC, BCH, Polar, Turbo), with particularly striking gains for longer codes and higher code rates, regimes in which existing neural approaches typically struggle.

\item \textit{Reproducibility:} We provide a complete implementation of TransCoder as a publicly available code \footnote{The source code is available at \url{https://github.com/smeshk/TransCoder}.} to enable further research and practical applications.

\end{itemize}

The remainder of the paper is organized as follows. Section \ref{sec:system_model} describes the communication system and defines the parts of the conventional transmission pipeline, including error-correction codes. Section \ref{sec:transcoder} presents the proposed TransCoder framework, i.e., the structure of neural modules and the organization of the information flow between the conventional and neural decoders. Section \ref{sec:num_results} provides numerical comparisons with a baseline without TransCoder and the SoTA neural decoders to evaluate the performance of the proposed framework. Finally, Section \ref{sec:conclusion} concludes this paper, while further details on the model implementation are provided in the Appendix \ref{sec:appendix}.

\textit{Notations:} Throughout this work, we use the following notations: $\bm{x} = [ x_0, x_1, \dots, x_{n-1}]$ refers to an $n$-dimensional vector, $x_i$ denotes the $i^\mathrm{th}$ element, a sequence of $n$-dimensional vectors refers to $X=[\bm{x}_0,\ldots,\bm{x}_{k-1}]$ or to a matrix form $\bm{X}_{k\times n}$ of shape $k\times n$. We use calligraphic $\mathcal{X}$ to denote an ordered set of elements $\left\{x_{0}, \ldots, x_{n-1}\right\}$, and use $\mathcal{Q}_{m}$ to refer to a particular ordered set  $\left\{\bm{q}_{0}, \ldots, \bm{q}_{2^m-1}\right\}$ of $m$-digit binary numbers, where $\bm{q}_{i}$ is the binary representation of integer $i$, for $i=0,\ldots,2^m-1$.

\section{System Model} \label{sec:system_model}

We consider a point-to-point additive white Gaussian noise (AWGN) channel as it is common in the literature, though the presented method can be directly applied to other channel models. Given a binary message $\bm{b}$ of length $k$, the goal of the transmitter is to convey this bit sequence reliably to the receiver over the AWGN channel. This is achieved by introducing redundancy into the sequence and encoding it into a new binary sequence of length $n > k$ to facilitate transmission over the noisy channel, that means employing a channel code $\mathcal{C}(n, k)$ defined over the binary field $\mathbb{F}_2$.\\
\indent Let this encoding operation be denoted by $E_{C(n, k)}$: $\bm{b}\in \left\{0,1\right\}^{k} \xrightarrow{} \bm{c} \in \{0, 1\}^n$, where codeword $\bm{c} \in \mathcal{C}(n, k)$ is the binary sequence encoded with the channel code $\mathcal{C}(n, k)$. The code rate of this encoding scheme is defined as $R = \frac{k}{n}$. In the scope of this work, we limit our focus to binary linear block codes  as they are mostly presented in the wireless transmission standards  \cite{3gpp}.\\ 
\indent After generating codewords, the transmitter maps each codeword $\bm{c}$ into a sequence of symbols $\bm{s}$ from the chosen constellation, i.e., for binary phase-shift keying (BPSK) modulation $\bm{s} \in \{-1, 1\}^n$, to be transmitted over the channel. We note that, in the most general case, the length of $\bm{c}$ need not be equal to that of $\bm{s}$ if larger constellations are used, e.g., QAM. We will restrict our presentation to BPSK modulation for simplicity.

The sequence of modulated symbols $\bm{s}$ is transmitted over the channel:
\begin{equation}
    \bm{y} = \bm{s} + \bm{w}\ , \label{eq:channel}
\end{equation}
where $\bm{w}$ represents the Gaussian noise component independent and identically distributed (i.i.d.) with $w_i \sim \mathcal{N}(0, \sigma^2), \ i=0,\dots,n-1$. An average power constraint, imposed on the symbol sequence $\bm{s}$, is expressed as follows:
\begin{equation}
\mathbb{E}_{\bm{b}} \left[ \frac{1}{n} \langle \bm{s}, \bm{s} \rangle \right] \leq 1\ , \label{avpov_exp}
\end{equation}
and is averaged over all information sequences $\bm{b}$, while for the BPSK case, the power is precisely equal to $1$. The signal-to-noise ratio (SNR) of this system is $\text{SNR} \triangleq 1 / \sigma^2$. At the receiver, the noisy sequence $\bm{y}$ is processed by the demodulator (if modulation was applied) to obtain the probabilities for every bit position to pass them to the channel decoder $D_{\mathcal{C}(n, k)}$.\\

\indent A common practice in implementing channel decoders \cite{3gpp} is to take the log likelihood ratios (LLRs) $\bm{l} \in \mathbb{R}^n$ as an input, where each LLR $l_i$ at position $i$,  $i=0,\dots,n-1$, is given by:
\begin{equation}
    l_i = \log \frac{\Pr(c_i=0|\bm{y})}{\Pr(c_i=1|\bm{y})}\ . \label{eq:llrs}
\end{equation}
For example, for the BPSK case, LLR computation can be simply expressed as:
\begin{equation*}
    l_i = \frac{y_i}{2\sigma^2} \ , \ \forall i = 0,\dots,n-1.
\end{equation*}
Having received the LLRs, channel decoder $D_{\mathcal{C}(n, k)}$ produces an estimate $\hat{\bm{b}}$ of the original message $\bm{b}$, i.e., $D_{\mathcal{C}(n, k)}$: $\bm{l}\in \mathbb{R}^n \xrightarrow{} \hat{\bm{b}} \in \{0, 1\}^k$.

The system performance can be evaluated by two key metrics: the block error rate (BLER) and the bit error rate (BER). BLER is defined as: 
\begin{equation}
        \BLER \triangleq \mathbb{E}_{\bm{b},\bm{w}} [\mathbbm{1}_{\left\{ \hat{\bm{b}} \neq \bm{b}\right\}}] \ , \label{eq:bler}
    \end{equation}
which quantifies the rate of incorrectly decoded codewords averaged over all possible transmitted information sequences $\bm{b}$ and all channel realizations $\bm{w}$. The BER is given by:
\begin{equation}
        \BER \triangleq \mathbb{E}_{\bm{b},\bm{w}} \left[\frac{1}{k}\sum^{k}_{i=1}\mathbbm{1}_{\left\{ \hat{b}_{i} \neq b_{i}\right\}}\right], \label{eq:ber}
    \end{equation}
representing the average rate of bit errors across all $k$ bits.

\subsection{Linear Block Codes}
A linear block code $\mathcal{C}(n, k)$ over the binary field $\mathbb{F}^n_2$ can be defined by the generator and parity-check binary matrix pair $\bm{G}_{k \times n}$ and $\bm{H}_{(n-k) \times n}$, respectively. The generator matrix $\bm{G}$ generates the codeword $\bm{c}$ by computing the parity checks from the information sequence $\bm{b}$, i.e.,
\begin{equation}
E_{\mathcal{C}(n,k)}: \bm{b}\in \left\{0,1\right\}^{k} \xrightarrow{} \bm{c} = \bm{b}\bm{G} \in \left\{0,1\right\}^{n} \ , \label{eq:channel_encoder}
\end{equation}
where the multiplication is performed over the binary field $\mathbb{F}_2$. The parity-check matrix $\bm{H}$ of size $(n-k) \times n$ satisfies $\bm{GH}^T =\bm{0}$ and $\bm{cH}^T = \bm{0}$ for any codeword $\bm{c} \in \mathcal{C}(n,k)$. For any given binary sequence $\hat{\bm{c}}\in \left\{0,1\right\}^{n}$, the parity-check matrix $\bm{H}$ can be used for syndrome calculation $\hat{\bm{c}}\bm{H}^T$, which is non-zero iff $\hat{\bm{c}} \notin \mathcal{C}(n,k)$. The parity-check matrix $\bm{H}$ is commonly used in decoding; for instance, in syndrome-based error correction or in BP decoding \cite{LDPC} performed over the Tanner graph constructed from $\bm{H}$. 

There are different implementations of linear block codes based on the specific properties/constructions of matrices $\bm{G}$ and $\bm{H}$. For example, polar codes~\cite{polar} perform channel polarization via a special construction of $\bm{G}$ and low density parity check (LDPC) codes \cite{LDPC} have a low density parity-check matrix $\bm{H}$. The choice of the linear code and its decoding algorithm relies on the system requirements. Depending on the decoding algorithm, the decoder returns a hard (binary) estimate of the codeword $\hat{\bm{c}}\in \left\{0,1\right\}^{n}$, for instance, successive cancellation (SC) decoding for polar codes, or a soft estimate of the codeword ${\bm{x}}\in \mathbb{R}^n$ that also can be treated as the {\em bit reliability estimates}, like in BP decoding for LDPC codes. To convert the soft estimate into a hard estimate, one can use the sign function, followed by a mapping to the binary domain, i.e., $\bm{\hat{c}} = 0.5 - 0.5\sign(\bm{x})$. In the scope of this work, we utilize soft estimate to enable iterative decoding for refinement.\\ 
\indent Next, we present the proposed TransCoder framework, which operates as an addition to the conventional channel encoder and decoder through transformer-based neural modules applied both at the transmitter and receiver.

\section{TransCoder}
\label{sec:transcoder}
The key idea behind the TransCoder design is to enhance the performance of a given channel code through the integration of additional neural encoder and decoder modules into the standard coding framework. Hence, as illustrated in Fig. \ref{fig:transcoder}, the overall architecture consists of five main blocks: two at the transmitter side, namely, channel encoder,  TransCoder encoder, and three at the receiver side, namely, TransCoder decoder, channel decoder, and TransCoder refinement decoder. Nevertheless, the TransCoder framework is adaptable to the system requirements and can be partially used in the transmission scheme. The limited integration of modules, i.e., only TransCoder encoder or decoder parts, does not change the idea of neural-assisted improvement of codeword weight distribution as illustrated in Fig.~\ref{fig:pairwise_distances}, though the whole pipeline brings the full benefits of the proposed scheme. Each of the TransCoder modules is based on the transformer encoder architecture with minor differences. Here, we will further explain these modules operating within the encoding/decoding phases separately.

\subsection{Encoding phase}
The encoding process at the transmitter consists of two stages. First, the binary information sequence $\bm{b}$ of length $k$ is passed through the channel encoder, producing the codeword $\bm{c}$ for the chosen channel code $\mathcal{C}(n, k)$, as in Eq. \ref{eq:channel_encoder}. Then, the codeword $\bm{c}$ is further processed by the TransCoder encoder $E_T$, which outputs a sequence of $n$ real symbols to be transmitted over the channel:
\begin{equation*}
    E_{T}(\cdot; \bm{\theta}): {\bm{c}}\in \left\{0,1\right\}^{n} \xrightarrow{} \bm{s}\in \mathbb{R}^n ,
\end{equation*}
where $\bm{\theta}$ are the trainable model parameters, and the vector $\bm{s}$ can also be denoted as a sequence of real symbols $\left[ s_{0}, \ldots, s_{n-1}\right], \ s_{i}\in \mathbb{R}, \ i=0,\dots,n-1$. To ensure the average power constraint from Eq.~(\ref{avpov_exp}) is satisfied, the TransCoder encoder has the power normalization module at the last stage consisting of symbol normalization and power reallocation, for which we will provide more details below.\\

\indent TransCoder encoder $E_T$ first exposes the codeword to a BPSK-based pre-modulation
$\bar{\bm{s}}= 2 \bm{c} - 1$, $\bar{\bm{s}}\in \left\{-1,1\right\}^{n}$. Then, the symbol vector $\bar{\bm{s}}$ is first divided into $n_b$ blocks of size $m$ each, to form a sequence ${S}_{bl} = \left[ \tilde{\bm{s}}_{0}, \ldots, \tilde{\bm{s}}_{n_{b}-1}\right]$, where $\tilde{\bm{s}}_{i}=[\bar{{s}}_{im},\dots,\bar{{s}}_{(i+1)m-1}]$, $i=0,\ldots,n_{b}-1$. The obtained sequence of blocks ${S}_{bl}$ is mapped by a block attention module (that follows a transformer encoder architecture) into a new sequence of $n_b$ blocks of the same size $m$, and it is further processed by the normalization module, as mentioned before.

\subsection{Decoding phase}
At the receiver side, complementary to the TransCoder encoder, the TransCoder decoder first divides the received noisy channel output $\bm{y}=[y_{0}, \ldots, y_{n-1}]$ into $n_b$ blocks of size $m$ each, ${Y}=  \left[ \tilde{\bm{y}}_{0}, \ldots, \tilde{\bm{y}}_{n_{b}-1}\right]$, where $\tilde{\bm{y}}_{i}=[y_{im}, \ldots, y_{(i+1)m-1}]$. Then, again, following the transformer encoding pipeline, each symbol block $\tilde{\bm{y}}_{i}$ is mapped to a $2^{m}$-dimensional probability vector $\tilde{\bm{p}}_{i}$, since the cardinality for the prediction  of each block is $2^{m}$; formally,

\begin{equation*}
D_{T}(\cdot; \bm{\theta}): {Y}\xrightarrow{}{P}_{bl} \ ,
\end{equation*}
where $\bm{\theta}$ are the trainable model parameters.

\indent The block-wise probabilities ${P}_{bl}= \left[ \tilde{\bm{p}}_{0}, \ldots, \tilde{\bm{p}}_{n_{b}-1}\right]$ computed for each symbol block can be converted to bit-wise LLRs for each of the $m$ bits in the block, using a message passing function $f_{m2d}({P}_{bl},m)$, as LLRs are required by the channel decoder. First, the probability estimates for all the bit positions are obtained for each block, i.e.,
\begin{equation}
{\bm{p}}_{i}=\sum^{2^{m}-1}_{j=0} \tilde{{p}}_{i,j} \bm{q}_{j}, ~~~~~\bm{q}_{j}\in \mathcal{Q}_{m} \ , \label{eq:per_bit_prob}
\end{equation}
where $\mathcal{Q}_{m}$ refers to the set of binary vector representations of integers from $0$ to $2^m-1$ and $\tilde{{p}}_{i,j}$ is $j$th element of the vector of probabilities $\tilde{\bm{p}}_{i}$ for $i$th block. Then, the sequence of bit-wise probabilities computed for each block $\left[\bm{p}_0,\dots,\bm{p}_{n_b - 1}\right]$ is flattened into a vector $\bm{p}$ of length $n$, and the function $f_{m2d}({P}_{bl},m)$ returns the bit-wise LLR estimates $\bm{l}$ utilizing the bit-wise probability estimates $p_i$, as the probability of having $1$ at the $i$th position, thus:
\begin{equation}
    l_i \triangleq \log \frac{1 - p_i}{p_i} \ , \label{eq:llrs_p}
\end{equation}
according to the LLR definition in Eq.~\ref{eq:llrs}.\\
\indent At the next stage, computed LLRs $\bm{l}$ are passed to the channel decoder $D_{\mathcal{C}(n,k)}$, which produces the message estimate $\hat{\bm{b}}$. If iterative refinement is enabled, the channel decoder outputs the soft codeword estimate ${\bm{x}}$, which we also refer to as the bit reliability estimates that are further used in the decoding pipeline.

\textbf{Iterative Refinement: } Iterative channel decoders utilize the bit reliability estimates ${\bm{x}}$ at each iteration as input to the next decoding iteration to improve their reliability. Extending this idea, TransCoder refinement decoder $D^{rf}_{T}$ aims to produce refined block-wise probability estimates, which are then fed back to the channel decoder. To exploit both the information from the noisy channel and the channel decoding results, the TransCoder refinement decoder $D^{rf}_{T}$ takes as input both the channel output and the bit reliability estimates ${\bm{x}}$ from the channel decoder. Then the refinement iterations can be performed in a loop. Thus, refinement decoder facilitates the partitioning of the channel decoding process and enhances error-correction capabilities through different decoding stages.

The TransCoder refinement decoder $D^{rf}_{T}$ has an identical structure to the TransCoder decoder $D_T$ with the only difference in the doubled input block length $2m$. During the $j$th refinement iteration, $j=1,\dots,N_{rf}$, a concatenation of channel outputs and the reliability estimates from the channel decoder is processed by $D^{rf}_{T}$ to generate the refined block-wise probabilities:
\begin{equation*}
D^{rf}_{T}(\cdot; \bm{\theta}): {Y}, {X}^{j}\xrightarrow{}{P}^{rf,j}_{bl}\ ,
\end{equation*}
where ${X}^{j}$ is a $j$th sequence of the bit reliability estimates produced by the channel decoder and processed by $f_{d2m}(\cdot)$ function before passing it to the TransCoder module. This function rescales the soft estimates $\bm{x}^{j}$ through inversing the LLR calculations and mapping the results into the $[-1,1]$ interval to align the values with channel outputs $\bm{y}$. To formalize, $f_{d2m}(\cdot)$ first applies the inverse of function in (\ref{eq:llrs_p}), and then applies BPSK pre-mapping:
\begin{equation}
    \hat{\bm{x}}^{j} = f_{d2m}(\bm{x}^{j}) = 2 \frac{\exp{\bm{x}^{j}}}{1 + \exp{\bm{x}^{j}}} - 1 \ . \nonumber
\end{equation}
The rescaled reliability estimates for the $j$th refinement iteration are given by ${X}^{j} = \left[ \tilde{\bm{x}}^{j}_{{0}}, \ldots, \tilde{\bm{x}}^{j}_{{n_{b}-1}}\right]$, where $\tilde{\bm{x}}^{j}_{{i}}=[{\hat{x}^{j}_{{im}}},\dots,{\hat{x}^{j}_{{(i+1)m-1}}}]$.

After the final refinement iteration, the channel decoder produces its final estimate of the transmitted information sequence $\bm{\hat{b}}\in \{0,1\}^k$.

\subsection{Architectural Details} 

\subsubsection{Block Attention Mechanism} \label{sec:bam}

As previously mentioned, proposed neural modules, TransCoder encoder $E_{T}$ and decoder $D_{T}$ (along with the refinement iteration decoder $D^{rf}_{T}$), share the same structure and are based on the transformer encoder architecture \cite{AttentionIsAllYouNeed} while also adopting a block-wise attention design \cite{gbaf, baaf}. As the TransCoder modules take the set of blocks of length $m$ as input, the vectors are padded first before reshaping if needed. For full architectural details, including pseudocode and component equations, see Appendix~\ref{sec:appendix_bam}.

\subsubsection{Power normalization and reallocation}
\label{main:power_constraint}

To ensure that the power constraint in Eq. (\ref{avpov_exp}) is satisfied on average, in our experiments, we consider the following approach. We assume that the symbol $s_{i}\in\mathbb{R}$ at the $i$th position of vector $\bm{s}$, $i = 0,\dots,n-1$, has a marginal distribution of $p_{s_{i}}$ with mean $\mu_{s_{i}}$ and variance $\sigma_{s_{i}}$. Then by normalizing symbols index-wise, i.e.,
\begin{equation}
    {s_{i}}' = \frac{s_{i} - \mu_{s_{i}}}{\sigma_{s_{i}}}, \label{eq:pw_norm}
\end{equation}
one can bound the average power consumption by $1$ for the $i$th symbol. By normalizing over all indices $i = 0,\dots,n-1$, one can approximately ensure the power constraint $(\ref{avpov_exp})$. However, there are two challenges to be addressed: first, it is difficult to predict the distribution $p_{s_{i}}$; second, these statistics vary during training. Hence, similarly to the well-known batch-normalization method used in training neural weights, we estimate these statistics over each batch during training in different channel conditions. Then, at test time, we take a significantly large batch, in the order of $10^{4}$, and compute the values of $\mu_{s_{i}}$ and $\sigma_{s_{i}}$ to collect the statistics for the given test setting and ensure the average power constraint for these conditions.

\label{main:power_realoc}
After symbol-wise normalization in Eq.~(\ref{eq:pw_norm}) described in the previous section, we multiply the sequence of averaged symbols ${\bm{s}}'$ by a learnable vector $\bm{w}$ of size $n$, $\vert\vert\bm{w}\vert\vert = 1$, which allows us to further optimize the power allocation over the transmitted symbols.

\subsection{Training Framework and Loss Function}
The training of the proposed neural TransCoder modules can be performed separately from channel decoding (without backpropagation through it) by optimizing the per bit probabilities $\bm{p}$ \footnote{In the block attention framework in \cite{baaf}, block-wise estimates in the form of multi-class predictions are employed; however, in our preliminary experiments, we observed that in our specific problem, where Transcoder is integrated with a channel decoder, bit-wise estimation, hence binary prediction, performs better.} given by TransCoder to the decoder in Eq.~(\ref{eq:per_bit_prob}), using binary cross-entropy loss for the target codeword $\bm{c}$:
\begin{equation}
    {\mathcal{L}_{TC} (\bm{p},\bm{c})} = \sum_{i=0}^{n} \{c_i \log(1 - p_i) + (1 - c_i) \log p_i \} \ . \label{eq:loss_tc}
\end{equation}

One of the major drawbacks of using the loss $\mathcal{L}_{TC}$ for training is that the learned TransCoder is oblivious to the channel decoder and targets solely minimizing the average bit-wise prediction error. However, this target may not be aligned with the correction capability of the channel decoder; that is, a probability distribution with a higher loss term according to Eq.~(\ref{eq:per_bit_prob}), does not necessarily imply a harder correction task for the channel decoder.\\
\indent In other words, we expect TransCoder to produce the LLRs from Eq.~(\ref{eq:llrs_p}) that are better aligned with the error correction capability of the particular channel decoder used. Hence, alternative to loss function $\mathcal{L}_{TC}$, we consider differentiable decoders and optimize the channel decoder predictions directly through end-to-end training. The loss term used for end-to-end training is also a  binary cross-entropy loss, where the true labels are bits of the target codeword $\bm{c}$ and the predictions are the  probabilities obtained by passing soft codeword estimates $\bm{x}$, given by the channel decoder, to sigmoid function $\sigma (\cdot)$:
\begin{multline}
    \mathcal{L}_{CD} (\bm{x},\bm{c}) = 
    \sum_{i=0}^{n} \{c_i \log(1 - \sigma ( x_{i})) + (1 - c_i) \log \sigma (x_{i}) \} \ .\label{eq:loss_cc}
\end{multline}
We do not consider any modifications to the integrated channel decoders in this work; therefore, the loss is taken at the end of channel decoding iterations/stage to allow for the optimization of TransCoder outputs for the specific channel decoding task. We used the binary cross-entropy loss after channel decoding for Turbo code and polar codes with an SC decoder, achieving the best results while taking the average of the losses after every iteration (every stage for the SC decoder), which is the loss used for training the neural decoders in \cite{tinyturbo, neural_polar_haim}.\\

\indent The binary cross-entropy optimizes the bit-wise probabilities or $\BER$ from Eq.~(\ref{eq:ber}), while the main metric of channel decoding is the probability of correct decoding or $\BLER$ from Eq.~(\ref{eq:bler}). Designing a loss function that optimizes $\BLER$ remains an open problem for channel decoders. Thus, for most neural channel decoders, the training is performed with binary cross-entropy loss. Though, for the Tanner graph-based BP decoder (for LDPC and BCH codes), we observed that training with only binary cross-entropy loss from Eq.~(\ref{eq:loss_cc}) shows suboptimal performance.\\
\indent In this work, we also introduce an additional loss term that functions as a regularizer for the model training. This approach tries to align the learning process with the code structure and is particularly efficient for BP decoding, which performs message passing based on the parity check matrix structure. \\ 
\textbf{Regularization with soft parity check values:}
\indent Before explaining the proposed regularization loss, we recall that during the conventional syndrome calculation, for the $i$th row of the parity check matrix $\bm{H}$, which forms a binary vector of length $n$, denoted by $\bm{h}_i = \bm{H}[i, :]$, $i = 0,\dots,n-k-1$, each syndrome value, i.e., the parity check, is computed modulo $2$ for the binary codeword estimate. Inspired by syndrome computation, we introduce the notion of {\em soft parity check}, which refers to the scalar vector product $(\bm{c},\bm{h}_i)$ for a vector $\bm{c}\in\mathbb{R}_{[0,1]}^n$. We remark that, from the construction, the soft parity check $(\bm{c},\bm{h}_i)$ for a codeword $\bm{c}\in\mathcal{C}$ can only take even values $0,2,\dots$ that are less than or equal to the parity check weight $|\bm{h}_i|$. The soft parity check can also be calculated for the soft codeword estimate $\bm{x}$ given by the channel decoder after projecting its values to the interval $[0,1]$:
\begin{equation}
    \bar{\bm{x}} = \frac{f_{cl}(\bm{x},[-1,1])+1}{2} \ , \label{eq:x_proj}
\end{equation}
where, $f_{cl}(\cdot,\cdot)$ is the clipping function. Then the soft parity check for the soft estimate $\bm{x}$ will be $(\bar{\bm{x}},\bm{h}_i)$ for $i$th parity check, while the target value is given with the soft parity check of the transmitted codeword $(\bm{c},\bm{h}_i)$.\\
\indent The underlying idea behind the use of the soft parity check values is to have a better distance metric between the codeword and its estimate by reformulating the problem as a multi-label classification problem. To be more precise, let $\mathcal{X}_{spc}$ be the set of all possible (and unique) soft parity check values for the given channel code with parity check matrix $\bm{H}$. We remark that $\mathcal{X}_{spc}$ is finite and its dimension $|\mathcal{X}_{spc}| \leq \max_{i} \floor*{|\bm{h}_i| /2}$.\\
\indent Since the set $\mathcal{X}_{spc}$ is finite, we can consider a soft parity check $(\bm{c},\bm{h}_i)\in \mathcal{X}_{spc}$ calculated for the codeword as a true label for each soft parity check value $(\bar{\bm{x}},\bm{h}_i)$. Hence, predicting a true label from soft parity check values $(\bar{\bm{x}},\bm{h}_i)$ can be considered as a multi-label classification problem. To be more precise, for a given $(\bar{\bm{x}},\bm{h}_i)$, we first evaluate its relation to all the labels from the set $\mathcal{X}_{spc}$ by computing the distances between the soft parity check value we obtained and all possible ones:
\begin{equation*}
    \bm{d}_{(\bar{\bm{x}},\bm{h}_i)} = \{|(\bar{\bm{x}},\bm{h}_i)-x|\}_{x\in \mathcal{X}_{spc}} \ .
\end{equation*}
Then, we  treat the obtained vector of distances  $\bm{d}_{(\bar{\bm{x}},\bm{h}_i)}$ as logit values, which can then be translated into log-probabilities through the log softmax $LSM(\cdot)$ layer. Hence, the final loss is categorical cross-entropy loss $CE(\cdot, \cdot)$ averaged over all $n-k$ soft parity checks calculated based on the soft codeword estimate $\bar{\bm{x}}$, i.e., 
\begin{multline}
    \mathcal{L}_{H} (\bar{\bm{x}},\bm{c}, \bm{H}) = \\ \frac{1}{n-k} \sum_{i=0}^{n-k-1} CE(LSM(\bm{d}_{(\bar{\bm{x}},\bm{h}_i)}),(\bm{c},\bm{h}_i)).\label{eq:loss_ldpc}
\end{multline}

When training with a BP channel decoder, we use a combination of two losses from Eqs.~(\ref{eq:loss_cc}) and (\ref{eq:loss_ldpc}) to ensure that both bit-wise and parity check-wise optimization is in place for the best results:
\begin{multline*}
     \mathcal{L}_{BP} ({\bm{x}},\bm{c}, \bm{H}) = \\ \frac{1}{2}\left( \mathcal{L}_{CD} (\bm{x},\bm{c}) +  \mathcal{L}_{H} \left(\frac{f_{cl}(\bm{x},[-1,1])+1}{2},\bm{c}, \bm{H}\right)\right) \ ,
\end{multline*}
according to the definition of projection in Eq.~(\ref{eq:x_proj}).

As the backpropagation process through multiple channel decoding stages might not provide accurate gradients, the loss for all considered channel codes was computed during every run of the channel decoder; thus, if there were $r-1$ refinement iterations, the loss would be computed $r$ times and averaged.

%%%%%%%%%%%%%%%%%%%%%%%%%%
\section{Numerical Results} \label{sec:num_results}

\begin{table*}[h]
    \centering
    \caption{Notations}
    \label{tab:notations}
{
\begin{tabular}{@{}ll@{}}
    \toprule
    Notation                    & Definition \\
    \midrule
    $\bm{b}$,    ${\bm{c}}$, $\bm{s}$                   & Bit  sequence,  encoded bits (codeword) by channel code, and encoded symbols by the neural encoder, respectively. \\
    $\bm{y}$,  $\bm{w}$,  $\bm{\hat{b}}$                   & Received noisy symbols, noise vector, estimate of $\bm{b}$ by channel decoder, respectively. \\
    $\bm{\hat{c}}$, $\bm{x}$              & Hard estimate of $\bm{c}$ by channel decoder and soft estimate of $\bm{c}$ (bit reliabilities), respectively.\\
    $k$, $n$                     & Length of the information bit sequence $\bm{b}$, total number of encoded symbols\\
    ${\mathcal{C}(n,k)}$ & Linear channel code with length $n$ and dimension $k$\\
    $E_{\mathcal{C}(n,k)}$,  $D_{\mathcal{C}(n,k)}$                       & Channel code encoder and decoder, respectively, for channel code $\mathcal{C}(n,k)$\\
    $E_T$, $D_T$ and $D^{rf}_T$  & TransCoder encoder, TransCoder decoder and TransCoder refinement iteration decoder, respectively\\
    BP-$i$,  NBP-$i$ & BP and neural BP (NBP) decoders with $i$ iterations  \\
    SC; SCL, L$=i$ & successive cancellation (SC) and successive cancellation list (SCL) with list size $i$ decoders (for polar codes)  \\

    \bottomrule
    \end{tabular}
    }
\end{table*}

In this section, we conduct extensive simulations to analyze the performance of the proposed TransCoder framework from the point of both computational complexity and the achieved error rate. We compare it with conventional code structures as well as the SoTA neural decoding algorithms. Table~\ref{tab:notations} provides the notations employed throughout this and previous sections.

\subsection{Simulation Setup}

\subsubsection{Code Structure} \label{sec:channel_codes_considered}
 For the numerical analysis, we consider a range of channel code families (taken from \cite{cc_database}) with different decoders and various code rates:
\begin{itemize}
    \item LDPC codes \cite{LDPC} with a BP decoder, which is part of the 5G NR standard \cite{3gpp}, and a neural modification of the BP decoder (NBP) \cite{neural_bp1} that introduces trainable weights to the graph over which the message passing algorithm is performed. We consider a range of code lengths from small $n=49$ to medium $n=121$, as well as longer codes $n=384$, compliant with the IEEE 802.22 wireless regional area network (WRAN) standard, with code rates varying from $0.5$ to $\sim0.83$. For the greater code length $n=384$, the NBP decoder is omitted due to the lack of any improvement.
    \item Polar codes \cite{polar} with SC and SCL \cite{scl_polar} decoders, while the latest is included in the 5G NR standard \cite{3gpp} (though we need to mention here that we use the SCL decoder with metric path calculation, whereas in the standard, an additional CRC code is used). We consider short $n=128$ and long $n=512$ codes with code rates varying from $0.5$ to $0.75$.
    \item BCH codes \cite{bch_1, bch_2}. We consider both the standard BP decoder and neural BP (NBP) decoder for code lengths $n=31$ and $n=63$, with code rates ranging from $~0.5$ to $~0.81$.
    \item Turbo codes \cite{turbo} used in the LTE standard with a turbo decoder. We consider a code length of $n=132$ with a code rate of $\sim0.3$.
\end{itemize}

For iterative decoders, we mention the number of iterations performed, as it drastically affects the decoding performance. For example, we denote the BP or NBP decoder with $i$ iterations as BP-$i$ or NBP-$i$, respectively. The number of turbo decoder iterations $i$ is denoted similarly as Turbo-$i$. For the SCL decoder for polar codes, the list size is equal to $8$ in our experiments, and we additionally denote this as $L=8$.

\subsubsection{Benchmarks}

 As the proposed framework includes both channel coding and neural modules, we provide as benchmarks both conventional channel decoders and SoTA neural decoders. Existing channel decoders offer adjustable complexity, improving performance by increasing the number of iterations or expanding the list size; thus, with a reasonably chosen complexity, they serve as natural baselines. For considered linear block codes, the SoTA neural decoding is transformer-based, and we limit our scope to the two most recent decoders, namely, AECCT \cite{aecct} and CrossMPT \cite{crossmpt}:
\begin{itemize}
    \item AECCT is an improved and accelerated version of the ECCT model \cite{ECCT}, the first fully neural transformer based decoder applicable to any linear block code. We report here only the parameters from the original paper \cite{aecct} that yield the best results, specifically the transformer encoder layers $N = 6$ and $N = 10$ with the embedding size $d=128$.
    \item CrossMPT is another transformer based neural decoder that shows improved performance compared to ECCT \cite{ECCT} by employing additional cross message passing during decoding, similarly to a conventional BP decoder. For this model, we also include the parameters that showed the best results in the original paper \cite{crossmpt}, which are $N = 6$ and $d=128$.
\end{itemize}
 For neural decoding benchmarks, which are transformer-based and show polynomial complexity growth with sequence length, we omit the results for longer block lengths, such as $n=512$ for polar codes\footnote{All the simulations were run on Nvidia RTX A6000 GPUs with 48GB of memory.}.

\subsubsection{TransCoder Configurations} \label{sec:configurations_transcoder}
In Section \ref{sec:transcoder}, we have shown that the proposed Transcoder framework has a modular structure. Based on the scenario and the performance target, TransCoder can be implemented with only a subset of the modules. Here we consider $5$ different scenarios of the proposed TransCoder framework integration, including the full pipeline:
\begin{itemize}
    \item $E_{T},\,D_{T},\,D^{rf}_{T}$ as the TransCoder encoder, decoder, and refinement decoder correspondingly;
\end{itemize}
and the variations with selected TransCoder modules integrated into the transmission scheme:
\begin{itemize}
    \item $E_{T},\,D_{T}$ as the TransCoder encoder and decoder, while the refinement decoder is omitted (for example, for non-iterative decoders like SC/SCL for polar codes);
    \item $D_{T},\,D^{rf}_{T}$ as the TransCoder decoder and refinement decoder, requiring only modifications at the receiver side;
    \item $D_{T}$ as the TransCoder decoder while the refinement decoder is omitted, again, requiring only modifications at the receiver side;
    \item $E_{T}$ as the TransCoder encoder, requiring only modifications at the transmitter side.
\end{itemize}

\begin{table*}[h]
    \centering
    \small
    \caption{Block error rate results $-\ln(\mathrm{BLER})$ for three different SNR values $E_b/N_0=\{4,5,6\} ~dB$ in each row. Higher values indicate better reliability (i.e., lower BLER). The best results are shown in \textbf{bold}, while the second best are \underline{underlined}. * Some values correspond to extremely low block error rates ($\mathrm{BLER}<10^{-8}$, or $-\ln(\mathrm{BLER})>18.5$), for which we could not collect reliable statistics due to an insufficient number of observed error events within our simulation limits.}
    \label{tab:main_table}
    \renewcommand{\arraystretch}{1.5}
    \resizebox{\textwidth}{!}{%
    % Total columns: 1 (Channel code) + 6 (Channel decoder) + 6 (ECCT) + 6 (AECCT) + 3 (CrossMPT) + 6 (TransCoder) = 28
    \begin{tabular}{l|cccccc|cccccc|ccc|cccccc}
    \toprule
    \makecell[l]{Channel code} 
      & \multicolumn{6}{c}{Channel decoder} 
      & \multicolumn{6}{c}{AECCT} 
      & \multicolumn{3}{c}{CrossMPT} 
      & \multicolumn{6}{c}{TransCoder} \\
    \midrule
     & \multicolumn{3}{c}{BP-40} & \multicolumn{3}{c}{NBP-40} 
     & \multicolumn{3}{c}{$N=6$} & \multicolumn{3}{c}{$N=10$} 
     & \multicolumn{3}{c}{$N=6$} 
     & \multicolumn{3}{c}{$D_{T},\,D^{rf}_T,\,4\times$BP-10} & \multicolumn{3}{c}{$E_{T},\,D_{T},\,D^{rf}_T,\,4\times$BP-10} \\
    \midrule
    %-------------- Segment 1: BP-40 --------------
    \makecell{\(\text{LDPC}\)\\\((49,24)\)}
      & 4.15 & 6.56 & 9.58 
      & 4.02 & 6.45 & 9.67
      & 3.71 & 6.22 & 9.50 
      & \(\underline{4.46}\) & \(\mathbf{7.28}\) & \(\mathbf{11.13}\)
      & \(\mathbf{4.50}\) & \(\underline{7.12}\) & \(\underline{10.76}\)
    & 4.12	 & 6.64 & 	9.95
    & 4.10	 & 6.59 & 	10.05 \\
      
    \makecell{\(\text{LDPC}\)\\\((121,60)\)}
      & \(\mathbf{3.27}\) & 6.29 & 10.80 
      & 3.25 & 6.50 & 11.51
      & 2.23 & 5.12 & 9.65 
      & 3.06 & 6.45 & 11.50
      & 3.14 & \(\underline{6.62}\) & \(\mathbf{11.86}\) 
      & \(\mathbf{3.27}\) & {6.60} & 11.50
      & \(\mathbf{3.27}\) & \(\mathbf{6.65}\) & \(\underline{11.70}\) \\
    % \midrule
    % %-------------- Segment 2: BP-30 --------------
    \makecell{\(\text{LDPC}\)\\\((121,70)\)}
      & 4.40 & 8.13 & 12.32
      & 4.42 & 8.18 & 12.09
      & 3.31 & 6.78 & 12.18
      & 4.30 & 8.20 & 13.24
      & 4.27 & \(\mathbf{8.58}\) & {13.40}
      & \(\mathbf{4.48}\) & 8.25 &  \(\mathbf{13.47}\)
      & \(\underline{4.43}\) & \(\underline{8.28}\) & \(\underline{13.43}\) \\
    \midrule
    %-------------- Segment 3: BP-30 --------------
     & \multicolumn{3}{c}{BP-30} & \multicolumn{3}{c}{NBP-30}
     & \multicolumn{3}{c}{$N=6$} & \multicolumn{3}{c}{$N=10$} 
     & \multicolumn{3}{c}{$N=6$} 
     & \multicolumn{3}{c}{\(D_{T},\,D^{rf}_{T}\,3\times\)BP-10} & \multicolumn{3}{c}{\(E_{T},\,D_{T},\,D^{rf}_T,\,3\times\)BP-10} \\
    \midrule
    \makecell{\(\text{LDPC}\)\\\((121,80)\)}
      & 5.07 & 8.93 & 14.07 
      & 5.15 & 9.09 & 14.10
      & 4.11 & 8.11 & 13.17
      & {5.10} & \(\mathbf{9.52}\) & {13.91}
      & 5.07 & \(\mathbf{9.52}\) & {13.91}
      & \(\mathbf{5.18}\) & 9.16 & \(\underline{14.24}\) 
      & \(\underline{5.16}\) & 9.27 & \(\mathbf{14.28}\) \\
    \midrule
    %-------------- Segment 4: BP-20 --------------
     & \multicolumn{3}{c}{BP-20} & \multicolumn{3}{c}{} 
     & \multicolumn{3}{c}{$N=6$} & \multicolumn{3}{c}{$N=10$} 
     & \multicolumn{3}{c}{$N=6$} 
     & \multicolumn{3}{c}{$D_{T},\,D^{rf}_T,\,2\times$BP-10} & \multicolumn{3}{c}{$E_{T},\,D_{T},\,D^{rf}_T,\,2\times$BP-10} \\
    \midrule
    \makecell{\(\text{LDPC}\)\\\((384,192)\)}
      & 10.00 & 11.66 & {N/A} 
      & --- & --- & --- 
      % & \multicolumn{3}{c}{}
      & 4.92 & 9.25 & {12.19}
      & 9.10 & 12.34 & 11.97
      & 6.79 & 11.48 & {12.35}
      & \(\mathbf{11.51}\) & \(\mathbf{15.65}\) &  \(\mathbf{>18.5}\)
      & \(\underline{10.90}\) & \(\underline{15.50}\) & \(\mathbf{>18.5}\) \\
      
    \makecell{\(\text{LDPC}\)\\\((384,256)\)}
      & 7.62 & 11.29 & 14.04 
      % & \multicolumn{3}{c}{}
      & --- & --- & --- 
      & 4.28 & 8.97 & 12.82 
      & 6.98 & 11.74 & {12.91}
      & 6.01 & 11.55 & {12.87}
      & \(\underline{7.71}\) & \(\underline{12.04}\) & \(\mathbf{>18.5}\)
      & \(\mathbf{8.07}\) & \(\mathbf{12.31}\) & \(\mathbf{>18.5}\) \\

    \makecell{\(\text{LDPC}\)\\\((384,320)\)}
      & 3.55 & 7.01 & 9.35 
      & --- & --- & --- 
      % & \multicolumn{3}{c}{}
      & 2.14 & 5.95 & 10.44 
      & 3.14 & \(\underline{7.59}\) & \(\mathbf{11.54}\) 
      & 2.96 & 7.50 & \(\underline{11.48}\) 
      & \(\underline{3.58}\) & 7.28 & 10.44 
      & \(\mathbf{3.76}\) & \(\mathbf{7.61}\) & 10.93 \\
    \midrule
    %-------------- Segment 6: BP-40 for BCH --------------
     & \multicolumn{3}{c}{BP-40} & \multicolumn{3}{c}{NBP-40} 
     & \multicolumn{3}{c}{$N=6$} & \multicolumn{3}{c}{$N=10$} 
     & \multicolumn{3}{c}{$N=6$} 
     & \multicolumn{3}{c}{$E_{T},\,BP-40$} & \multicolumn{3}{c}{$E_{T},\,D_{T},\,D^{rf}_T,\,8\times$BP-5} \\
    \midrule
    \makecell{\(\text{BCH}\)\\\((31,16)\)}
      & 2.64 & 3.82 & 5.25 
      & 2.72 & 4.11 & 5.91
      & \(\underline{4.87}\) & \(\underline{6.75}\) & \(\underline{9.60}\) 
      & \(\mathbf{5.37}\) & \(\mathbf{7.38}\) & \(\mathbf{10.55}\)
      & 3.98 & 5.94 & 8.56
      & 3.54 & 5.17 & 7.16 
      & 3.88 & 5.54 & 7.69 \\
      
    \makecell{\(\text{BCH}\)\\\((63,36)\)}
      & 1.30 & 2.22 & 3.51 
      & 1.21 & 2.07 & 3.24
      & 2.13 & 3.65 & 5.85 
      & 2.37 & 3.97 & 6.22
      & 2.20 & 3.98 & \(\underline{6.61}\)
      & \(\underline{2.65}\) & \(\underline{4.18}\) & 6.22
      & \(\mathbf{3.05}\) & \(\mathbf{4.97}\) & \(\mathbf{7.48}\) \\
      
    \makecell{\(\text{BCH}\)\\\((63,45)\)}
      & 2.08 & 3.57 & 5.30 
      & 2.03 & 3.69 & 5.92
      & \(\mathbf{3.03}\) & \(\mathbf{5.22}\) & \(\mathbf{8.32}\)
      & \(\underline{2.98}\) & \(\underline{5.08}\) & \(\underline{8.07}\)
      & 2.57 & 4.60 & 7.52
      & 2.75 & 4.58 & 6.83 
      & 2.85 & 4.90 & 7.68 \\
      
    \makecell{\(\text{BCH}\)\\\((63,51)\)}
      & 1.59 & 2.78 & 4.21 
      & 1.47 & 2.82 & 4.60
      & \(\underline{2.79}\) & 4.87 & 8.01 
      & \(\mathbf{3.01}\) & \(\mathbf{5.28}\) & \(\mathbf{8.48}\)
      & \(\underline{2.79}\) & \(\underline{5.03}\) & \(\underline{8.02}\)
      & 1.92 & 3.38 & 5.22 
      & 1.98 & 3.71 & 6.12 \\
    \midrule
    & \multicolumn{3}{c}{SC} & \multicolumn{3}{c}{SCL, $L=8$} 
     & \multicolumn{3}{c}{$N=6$} & \multicolumn{3}{c}{$N=10$} 
     & \multicolumn{3}{c}{$N=6$} 
     & \multicolumn{3}{c}{$E_{T},\,D_{T},\,SC$} & \multicolumn{3}{c}{$E_{T},\,D_{T},\,SCL,L=8$} \\
    \midrule
    \makecell{\(\text{Polar code}\)\\\((128,64)\)}
      & 5.52 & 8.20 & 11.35 
      & 5.78 & 8.43 & 11.50 
      & 1.50 & 3.19 & 5.62
      & 2.85 & 5.34 & 8.64
      & 2.50 & 6.62 & 11.86 
      & \(\underline{6.19}\) & \(\underline{9.29}\) & \(\underline{12.66}\) 
      & \(\mathbf{6.82}\) & \(\mathbf{9.60}\) & \(\mathbf{13.01}\) \\ 
    \makecell{\(\text{Polar code}\)\\\((128,86)\)}
      & 4.17 & 6.24 & 8.75 
      & {4.32} & 6.29 & 8.75
      & 2.11 & 4.24 & 7.44 %
      & 3.58 & 6.31 & 10.25 
      & 3.93 & 7.06 & 11.23 
      & \(\underline{4.95}\) & \(\underline{8.06}\) & \(\underline{12.03}\) 
      & \(\mathbf{6.61}\) & \(\mathbf{10.16}\) & \(\mathbf{14.29}\) \\
      \makecell{\(\text{Polar code}\)\\\((128,96)\)}
      & 3.67	& 5.80 & 8.46 
      & 3.98	& 6.05 & 8.74  
      & 2.27 & 4.66 & 7.82 
      & 3.13 & 5.97 & 9.00  
      & 3.52 & 6.32 & 9.18 
      & \(\underline{4.00}\) & \(\underline{6.71}\) & \(\underline{9.41}\) 
      & \(\mathbf{4.96}\) & \(\mathbf{7.26}\) & \(\mathbf{9.68}\) \\ 
      \makecell{\(\text{Polar code}\)\\\((512,256)\)}
      & 7.98	& 11.04 & 13.96 
      & 8.29	& 11.05 & 14.41 
      & --- & --- & --- 
      & --- & --- & --- 
      & --- & --- & --- 
      & \(\underline{9.79}\) & \(\underline{14.58}\) & \(\mathbf{19.33}\)
      & \(\mathbf{12.24}\) &  \(\mathbf{16.31}\) & \(\mathbf{>18.5}\) \\
      \makecell{\(\text{Polar code}\)\\\((512,384)\)}
      & 4.96	& 8.79 & 13.56 
      & \(\underline{6.28}\)	& \(\underline{10.17}\) & \(\underline{14.55}\)  
      & --- & --- & --- 
      & --- & --- & --- 
      & --- & --- & --- 
      & {5.16} & {9.39} & {14.10}  
      & \(\mathbf{6.81}\) & \(\mathbf{10.61}\) & \(\mathbf{15.31}\)  \\
      \midrule
    %-------------- Segment 10: Turbo-2 (Turbo LTE) --------------
     & \multicolumn{3}{c}{Turbo-2} & \multicolumn{3}{c}{} 
     & \multicolumn{3}{c}{$N=6$} & \multicolumn{3}{c}{$N=10$} 
     & \multicolumn{3}{c}{$N=6$} 
     & \multicolumn{3}{c}{$E_{T},\,D_{T},\,$Turbo-2} & \multicolumn{3}{c}{$E_{T},\,D_{T},\,D^{rf}_T,\,2\times$Turbo-1} \\
    \midrule
    \makecell{\(\text{Turbo LTE}\)\\\((132,40)\)}
      & 6.38 & 9.51 & 12.98 
      & --- & --- & --- 
      & 0.69 & 1.55 & 2.92 
      & 1.88 & 3.45 & 5.72 
      & 2.70 & 4.76 & 7.68
      & \(\underline{6.79}\) & \(\underline{9.77}\) & \(\underline{13.16}\)
      & \(\mathbf{7.22}\) & \(\mathbf{10.37}\) & \(\mathbf{13.49}\) \\
    \bottomrule
    \end{tabular}%
    }
  \end{table*}

We denote the number of channel decoder runs $r$ in the TransCoder framework ($r>1$ if iterative refinement decoding is enabled; with $r-1$ denoting the number of refinement iterations) as $r\text{x} D_{\mathcal{C}(n,k)}$. 

\subsubsection{Training framework}
The proposed end-to-end communication framework utilizing  TransCoder neural modules was trained in an end-to-end fashion with fixed and backpropagatable channel decoders. During the training, the average signal magnitude was fixed to $1$ and the SNR $E_b/N_0$ of the considered AWGN channel was uniformly sampled for every transmission from the interval $[ \,2,8] \,$ dB. Both training and testing datasets were generated by uniform sampling of a binary information bit sequence $\bm{b} \in \{0,1\}^{k}$ that is then encoded into a codeword $\bm{c} \in \{0,1\}^{n}$ with the underlying channel code $\mathcal{C}(n, k)$. We use both $\BER$ and $\BLER$ metrics to evaluate the scheme's performance, ensuring that there were at least $10^6$ transmitted codewords to collect enough statistics and at least $100$ messages were incorrectly decoded. The hyperparameters of the TransCoder modules and the training details can be found in Table~\ref{tab:model_param} in Appendix~\ref{appendix:implementation_details}.

\subsection{Experimental Results}

Considering the channel codes mentioned in Section \ref{sec:channel_codes_considered}, we present the performance results for $-{\ln(\text{BLER})}$ in Table~\ref{tab:main_table} for the channel decoding and neural-based benchmarks, along with the two different configurations of the proposed TransCoder framework defined in Section \ref{sec:configurations_transcoder}. The chosen TransCoder configurations demonstrate the performance of the full pipeline and the second best configuration, with some of the TransCoder modules omitted. The ablation study of the different TransCoder configurations is provided further in Section \ref{subsec:further_ablations}. For every considered setting, the performance is evaluated at $E_b/N_0 \in \{4, 5, 6\}$ dB, except in cases where we could not collect enough statistics during the simulation.

\begin{table*}[t!]
\centering
% \small
\caption{Comparison of the computational complexity of different models. ECCT results are shown first, followed by $E_T, D_T, D_T^{rf}$. Inference times are measured per batch ($bs=1000$). For ECCT, we report complexity results for a simpler version ($N=6$).}
\label{tab:complexity}
\renewcommand{\arraystretch}{1.1}
{
\begin{tabular}{lcccccc}
\toprule
& \multicolumn{2}{c}{\# Params} 
& \multicolumn{2}{c}{\# FLOPs} 
& \multicolumn{2}{c}{Inference Time} \\
\cmidrule(lr){2-3}
\cmidrule(lr){4-5}
\cmidrule(lr){6-7}
Method
& ECCT
& $E_T, D_T, D_T^{rf}$
& ECCT
& $E_T, D_T, D_T^{rf}$
& ECCT
& $E_T, D_T, D_T^{rf}$ \\
\midrule
LDPC (121,70)
& 1.2M 
& 10k, 15k, 15k
& 229.7M
& 514k, 764k, 770k
& 37.0ms
& 2.2ms, 2.6ms, 2.7ms \\

LDPC (384,320)
& 1.4M
& 10k, 15k, 15k
& 53.1M
& 1.5M, 2.3M, 2.3M
& N/A
& 2.0ms, 2.7ms, 2.8ms \\

BCH (63,45)
& 1.2M
& 10k, 15k, 15k
& 99.8M
& 236k, 351k, 354k
& 27.0ms
& 2.1ms, 2.6ms, 2.6ms \\
\bottomrule
\end{tabular}
}
\end{table*}

\subsubsection{LDPC codes} \label{sec:ldpc_res} We considered a range of LDPC codes, specifically code lengths $n = \{49, 121, 384\}$ and different code rates. For all LDPC codes, we provide the results for the full TransCoder pipeline and the scenario with integrated TransCoder decoders only, as it presents the best results for these codes. The comparison to the BP baseline shows that both considered TransCoder configurations significantly improve the BLER performance, especially for higher SNRs. Moreover, the NBP decoder does not demonstrate a similar order of improvement compared to the scenario with TransCoder decoders, showing the advantages of the proposed model. However, neural decoding benchmarks, AECCT $N=10$ and CrossMPT $N=6$, show the best performance for shorter codes, more specifically with $n=49$ and $n=121$, and higher code rates, close to $0.5$, at high SNR, while the TransCoder framework is beneficial for longer code lengths and lower code rates that might be challenging for neural decoders.

\begin{tcolorbox}
    \textbf{Takeaway \ref{sec:ldpc_res}:} TransCoder outperforms NBP decoders for LDPC codes for both full pipeline and configuration with only TransCoder decoders. 
\end{tcolorbox}

\subsubsection{BCH codes} \label{sec:bch_res}
As the BCH codes are considered with the BP decoder, which is suboptimal for this class of codes, the improvements shown by neural decoders and the TransCoder framework are more profound compared to LDPC codes with the BP decoder, although the TransCoder relies on channel decoding within the decoding pipeline. That is also why, for these codes, the second best TransCoder framework would be the $E_T$, which has only the integrated TransCoder encoder at the transmitter side, as it reshapes the codeword weight distribution, as illustrated in Fig.~\ref{fig:pairwise_distances}. Similarly to the LDPC codes, the TransCoder framework is beneficial for longer code length $n=63$ and lower code rates $\sim0.5$, while the transformer-based AECCT decoder outperforms other schemes for the shortest considered block length of $n=31$ and at higher code rates.

\begin{tcolorbox}
    \textbf{Takeaway \ref{sec:ldpc_res} and  \ref{sec:bch_res}:} For BCH codes with BP decoding, TransCoder shows the best performance gain with longer block lengths and lower code rates compared to other transformer-based approaches. 
\end{tcolorbox}

\subsubsection{Polar codes} \label{sec:polar_res}
For the considered polar codes, the TransCoder framework has shown significant improvement compared to SC and SCL decoders. Notably, for code length $n=128$ and for longer codes $(512, 256)$ at lower code rates the TransCoder with the SC decoder outperforms the conventional SCL decoder with path metric calculation for a list size of $L=8$. As SC and SCL decoders are not iterative, the TransCoder refinement decoding is not enabled in this case. Moreover, the neural decoders exhibit increasing complexity for longer codes, and we are unable to provide the results for code lengths $n=512$ due to model complexity. For a moderate code length of $n=128$, TransCoder, even with a simpler decoder, i.e., SC, outperforms the state-of-the-art neural decoders for all considered code rates. This can be explained by the fact that these models use masking based on the parity check matrix, which is less informative due to its high density for polar codes compared to LDPC codes. 

\subsubsection{Turbo codes}\label{sec:turbo_res}
The considered Turbo code $(132, 40)$ has the lowest code rate among the codes in our experiments, and in this case, the modern transformer based decoders drastically fail compared to the well-established methods, such as the turbo decoder. On the other hand, TransCoder improves the performance, both with and without refinement decoding enabled.

\begin{tcolorbox}
    \textbf{Takeaway \ref{sec:polar_res}:} For polar codes, TransCoder trained with the SC decoder improves the performance of both the SC and SCL decoders, and for most of the codes, the performance gain for the TransCoder pipeline with the SC decoder is greater compared to the improvement offered by SCL with path metric calculation. 
\end{tcolorbox}

\begin{tcolorbox}
    \textbf{Takeaway \ref{sec:ldpc_res}, \ref{sec:bch_res}, \ref{sec:polar_res} and \ref{sec:turbo_res}:} For all the code families that have been investigated, e.g., BCH, LDPC, Polar, TransCoder outperforms the channel decoding baseline, especially, in the high SNR regimes. 
\end{tcolorbox}

\subsection{Complexity Analysis}
\label{sec:complexity}
\begin{table*}[h]
\centering
\small
\caption{Comparison of the number of the multiplications or additions for different models.}
\label{tab:complexity_theory}
{
\begin{tabular}{lcccc}
\toprule
Method
& Part of the model
& Number of
& \multicolumn{2}{c}{Comparison for } \\

& 
& multiplications/additions
& LDPC $(121, 60)$ & LDPC $(384, 320)$  \\
% \cmidrule(lr){2-3}
% \cmidrule(lr){4-5}
% \cmidrule(lr){6-7}
% Method
% & BP
% & 1 iteration
% & ECCT
% & $E_T, D_T, D_T^{rf}$
% & ECCT
% & $E_T, D_T, D_T^{rf}$ \\
\midrule
BP
& $1$ iteration
& $E$
& $726$ & $1280$ \\

& $50$ iterations
& $50E$
& $36$k & $64$k \\
\hline
TransCoder
& encoder layer 
& $12n_{b}d_{model}^2 + n_{b}^2 d_{model} + 2n_{b}^2$
& 156k & 461k \\

& $FE(\cdot)$ and embedding 
& $12n_{b}d_{model}^2 + 3mn_{b} d_{model}$
& 132k & 313k \\

& $E_T$
& 2 layers+emb %$2(12n_{blocks}d_{model}^2 + n_{blocks}^2 d_{model} + 2n_{blocks}^2)+12n_{blocks}d_{model}^2 + 3mn_{blocks} d_{model}$
& 444k & 1.2M \\

& $D_T$ or $D^{rf}_T$
& 3 layers+emb
& 600k & 1.7M \\
\hline
CrossMPT
& encoder layer 
& $(11n+13n_{pc})d_{cm}^2 + 4 n_{pc} n (d_{cm}+h)$
% & $12n_{b}d_{model}^2 + n_{b}^2 d_{model} + 2n_{b}^2$
& 40M & 96M \\

& final embedding 
& $n (n + n_{pc})$
& 22k & 172k \\

& decoder
& 6 layers+emb
& 241M & 577M \\
\hline
ECCT
& encoder layer 
& $12(n+n_{pc})d_{cm}^2 + 2 (n_{pc} + n)^2 (d_{cm}+h)$
% & $12n_{b}d_{model}^2 + n_{b}^2 d_{model} + 2n_{b}^2$
& 46M & 143M \\

& final embedding 
& $(n + n_{pc})^2$
& 35k & 200k \\

& decoder
& 6 layers+emb
& 278M & 856M \\

% LDPC (384,320)
% & 1.4M
% & 10k, 15k, 15k
% & 53.1M
% & 1.5M, 2.3M, 2.3M
% & N/A
% & 2.0ms, 2.7ms, 2.8ms \\

% BCH (63,45)
% & 1.2M
% & 10k, 15k, 15k
% & 99.8M
% & 236k, 351k, 354k
% & 27.0ms
% & 2.1ms, 2.6ms, 2.6ms \\
\bottomrule
\end{tabular}
}
\end{table*}
TransCoder's practical utility hinges on its computational efficiency compared to standard BP and state-of-the-art transformer-based methods. Table~\ref{tab:complexity} provides a detailed comparison of computational requirements across different neural approaches.
Note that ECCT with $N=10$ can achieve comparable performance to our approach but requires substantially more computational resources than the $N=6$ baseline. 

Unlike fully transformer-based models that require extensive computational resources and memory, TransCoder's efficiency stems from three design choices: i) Block attention mechanism that processes data in manageable chunks rather than attending to entire sequences; ii) Modular architecture that allows selective deployment of components based on system requirements; iii) Efficient iterative refinement strategy that maximizes performance gain per computational cost. For example, as shown in Table \ref{tab:complexity}, TransCoder requires only $10-15$k parameters per module for LDPC $(121,70)$, compared to $1.2$M parameters for equivalent transformer-based approaches. This significant reduction in parameter count, combined with efficient block processing, makes TransCoder particularly suitable for practical deployments.

For further analysis of the theoretical (time) complexity of neural decoders compared to conventional decoders, we provide a number of multiplications/additions for the BP decoder, CrossMPT \cite{crossmpt}, and ECCT \cite{ECCT} decoders, and TransCoder modules in Table~\ref{tab:complexity_theory} for selected LDPC codes:
\begin{itemize}
    \item  For the BP algorithm over a Tanner graph defined for LDPC codes in \cite{bp_standart}, the number of multiplications/additions in one BP iteration is equal to the number of edges in the Tanner graph, and it has the lowest complexity.
    \item  The CrossMPT and ECCT \footnote{In Table \ref{tab:main_table}, the performance results are provided for AECCT, an improved ECCT model. Given that the architectures are similar, we report here the complexity of the ECCT model, which does not use any quantization and is straightforward for complexity evaluation.} decoders are transformer-based and have similar architectures; they also show similar complexity, which is the amount of operations in the attention layers. The complexity results here are given for transformer encoder layers $N=6$, embedding size $d=128$, the number of attention heads $h=8$. We can see that the complexity of these decoders exceeds that of conventional decoders by more than $4$ orders of magnitude.
    \item TransCoder modules are also transformer-based; however, the use of block attention with a number of blocks $n_{b} = \lceil \frac{n}{m} \rceil$ and a much smaller embedding size $d_{model} = 16$ results in a significant decrease in complexity. The block size is $m=3$ for LDPC $(121,60)$ and $m=4$ for LDPC $(384,320)$. Results show that the complexity of the TransCoder module is more than $2$ orders of magnitude less compared to ECCT.
\end{itemize}

\begin{figure}[t]
    \centering
    \includegraphics[width=.9\columnwidth]{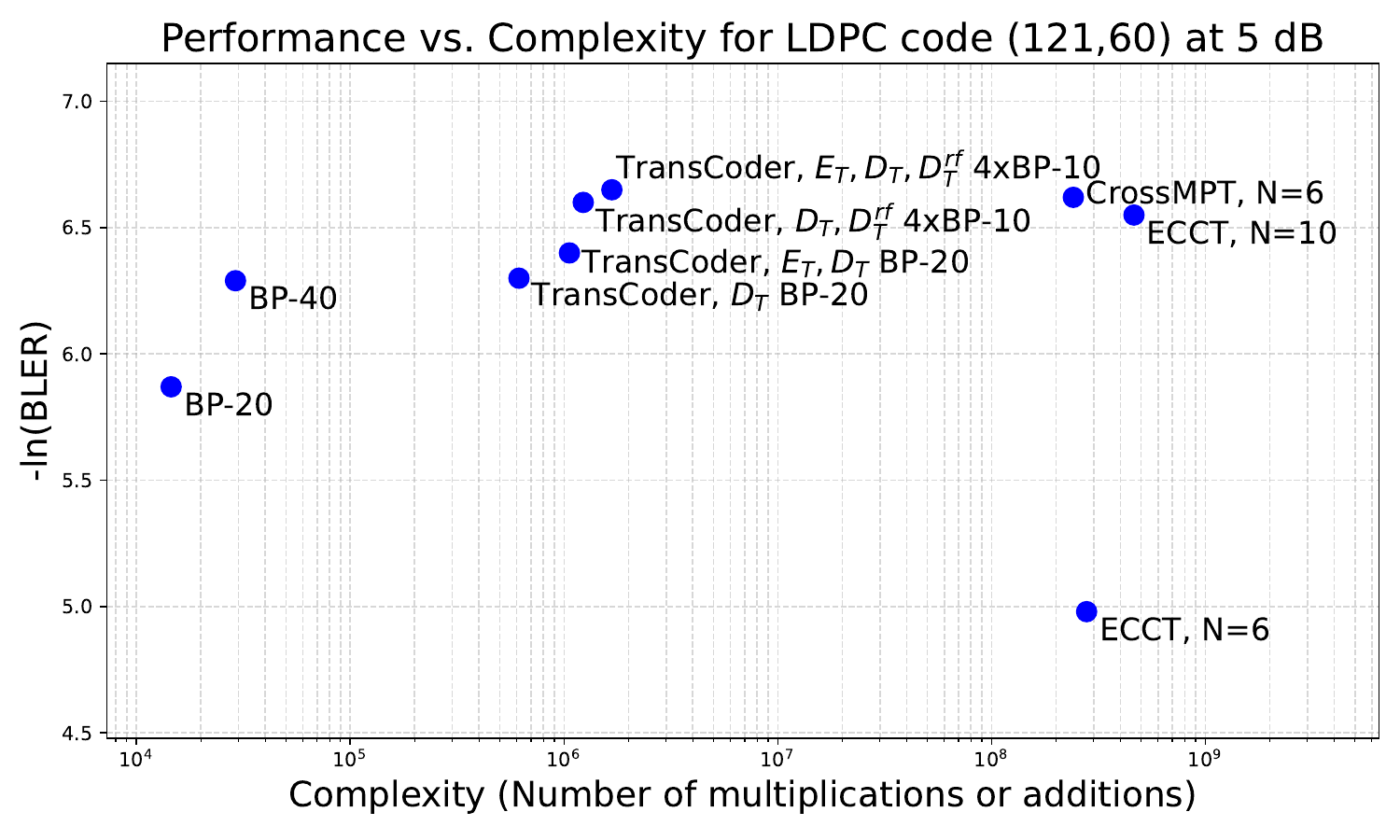}
    \caption{Performance as $-\ln(BLER)$ vs. complexity as number of multiplications or additions for LDPC code $(121, 60)$ at $5$~dB.}
    \label{fig:perfvscompl}
\end{figure}

The proposed TransCoder framework achieves SoTA neural decoding performance while maintaining a concise model architecture, as illustrated in Fig.~\ref{fig:perfvscompl} for LDPC code $(121, 60)$ at $5$~dB. While large-scale transformer baselines like CrossMPT and ECCT reach high performance peaks, they incur astronomical computational costs. In contrast, the TransCoder framework leverages underlying channel decoding to bridge the gap between traditional BP and heavy neural models. The most advanced TransCoder configuration $(E_{T},\,D_{T},\,D^{rf}_{T})$  achieves superior performance, outperforming CrossMPT, while operating at a lower complexity. This represents a complexity reduction of over two orders of magnitude compared to pure transformer-based approaches, demonstrating that the TransCoder framework offers a highly efficient trade-off between neural capability and computational tractability.

\subsection{Ablation Study on TransCoder Configurations} 
\label{subsec:further_ablations}

\begin{table*}
\centering
\small 
\caption{Comparison of $-\ln(\mathrm{BLER})$ for three different SNR values $E_b/N_0 = \{4,5,6\} \ dB$ in each row. Higher values indicate better reliability (i.e., lower BLER). The best results are shown in \textbf{bold}, while the second best are \underline{underlined}.}
\label{tab:ablation_table}
\renewcommand{\arraystretch}{1.0}
{
    \begin{tabular}{cccc|ccc|ccc|ccc|ccc|ccc}
  \toprule
   & \multicolumn{3}{c}{ Channel decoder} & \multicolumn{12}{c}{TransCoder} \\

   Channel & \multicolumn{3}{c}{ \ } & \multicolumn{3}{c}{$D_{T}$} & \multicolumn{3}{c}{$D_{T}$, $D^{rf}_T$} & \multicolumn{3}{c}{$E_{T}$, $D_{T}$} & \multicolumn{3}{c}{$E_{T}$, $D_{T}$, $D^{rf}_T$} \\
   \makecell[l]{code}  & \multicolumn{3}{c}{ BP-20 } & \multicolumn{3}{c}{BP-20} & \multicolumn{3}{c}{$2\times$BP-10} & \multicolumn{3}{c}{BP-20} & \multicolumn{3}{c}{$2\times$BP-10} \\
  \midrule
  \makecell{$\text{LDPC}$ \\ ${(49, 24)}$ } & {3.93} & 6.15 & 9.14 & 3.87 & {6.30} & 9.46 & \textbf{3.97} & \textbf{6.41} & {9.53} & 3.86 & 6.25 & \underline{9.54} & \underline{3.95} & \underline{6.37} & \textbf{9.67}  \\ 
  \midrule
  \makecell{$\text{LDPC}$ \\ ${(121, 60)}$ } & 3.09 & 5.87 & 9.84 & \underline{3.14} & 6.33 & 10.76 & \textbf{3.15} & 6.32 & 10.66 & 3.13 & \textbf{6.40} & \underline{11.15} & \underline{3.14} & \underline{6.37} & \textbf{11.17} \\
  \midrule
  \makecell{$\text{LDPC}$ \\ ${(121, 70)}$ }  & 4.09 & 7.37 & 11.92 & \underline{4.24} & \underline{7.85} & 12.35 & \underline{4.24} & 7.77 & 12.09 & 4.23 & \textbf{7.88} & \underline{12.65} & \textbf{4.26} & 7.82 & \textbf{12.76} \\
  \midrule
   & \multicolumn{3}{c}{ \ } & \multicolumn{3}{c}{$E_{T}$} & \multicolumn{3}{c}{$D_{T}$, $D^{rf}_T$} & \multicolumn{3}{c}{$E_{T}$, $D_{T}$} & \multicolumn{3}{c}{$E_{T}$, $D_{T}$, $D^{rf}_T$} \\
  & \multicolumn{3}{c}{ BP-20 } & \multicolumn{3}{c}{BP-20} & \multicolumn{3}{c}{$4\times$BP-5} & \multicolumn{3}{c}{BP-20} & \multicolumn{3}{c}{$4\times$BP-5} \\
  \midrule
  \makecell{$\text{BCH}$ \\ ${(31,16)}$  \\} & 2.45 & 3.43 & 4.62 & 3.39 & 4.86 & 6.71 & 2.59 & 3.79 & 5.37 & \underline{3.60} & \underline{5.21} & \underline{7.22} & \textbf{3.83} & \textbf{5.50} & \textbf{7.65} \\ 
  \midrule
  \makecell{$\text{BCH}$ \\ ${(63,36)}$  \\} & 1.25 & 2.06 & 3.17 &  2.53 & 3.92 & 5.69 & 1.28 & 2.34 & 3.92 & \underline{2.72} & \underline{4.46} & \underline{6.52} & \textbf{2.98} & \textbf{4.89} & \textbf{7.39} \\ 
    \midrule
  \makecell{$\text{BCH}$ \\ ${(63,45)}$  \\}  & 2.01 & 3.28 & 4.87 & \underline{2.64} & 4.32 & 6.36 & 2.02 & 3.71 & 6.01 & 2.61 & \underline{4.50} & \underline{6.82} & \textbf{2.80} & \textbf{4.81} & \textbf{7.61} \\
  \midrule
  \makecell{$\text{BCH}$ \\ ${(63,51)}$  \\}  & 1.56 & 2.69 & 4.07 & \underline{1.89} & \underline{3.29} & 5.08 & 1.50 & 2.89 & 4.99 & 1.77 & 3.28 & \underline{5.34} & \textbf{1.93} & \textbf{3.63} & \textbf{6.01} \\
  \bottomrule
\end{tabular}
}
\end{table*}

Next, we investigate the effect of the TransCoder configurations further. In Table~\ref{tab:ablation_table}, we compare the performance in $-{\ln(\text{BLER})}$ while integrating different TransCoder modules into the transmission scheme against baseline BP-$20$ decoding at $E_b/N_0 = \{4, 5, 6\}$ dB across both LDPC and BCH families. 

For LDPC codes, we omitted the configuration with only the TransCoder encoder $E_T$ module, as it does not show significant improvement for these codes. The two configurations with only decoders employed, $D_T$ and $D_{T}, D_{T}^{rf}$, show particularly significant improvements in lower SNRs. The best improvements in the high SNR regime, however, are demonstrated by the full TransCoder pipeline $E_{T}, D_{T}, D_{T}^{rf}$ for all considered LDPC codes.

For BCH codes, the performance boost provided by the TransCoder encoder $E_{T}$ is significantly greater than the performance gains from the configuration with TransCoder decoders $D_{T}, D_{T}^{rf}$. Thus, we observe that for different code families, different configuration choices demonstrate the most significant improvements.

\section{Conclusion} \label{sec:conclusion}
We introduced TransCoder, a novel neural performance enhancing transmission framework that combines the benefits of the existing structure of error correction codes with the power of modern transformer architectures. Our approach uniquely combines three key innovations: adaptive neural encoding at the transmitter, block-attention processing for efficient decoding, and an iterative refinement mechanism that enhances decoding accuracy. Through extensive experimentation across major code families (LDPC, BCH, Polar, and Turbo), we have demonstrated that TransCoder achieves significant improvements in block error rate (BLER) while maintaining computational efficiency comparable to traditional BP and other decoders. TransCoder's modular design enables flexible deployment at either the transmitter or receiver, making it particularly valuable for practical systems with varying computational and latency constraints. e.g., the decoder can dynamically decide on the number of decoding iterations depending on the available computational resources and the delay/reliability trade-off required for the underlying application. It also has the advantage of benefiting from efficient hardware implementations of legacy encoding and decoding algorithms. By reshaping the distribution of codeword distances without modifying the underlying code structures, TransCoder establishes a new paradigm for error correction that bridges conventional coding theory and neural approaches, offering a practical path forward for next-generation communication systems.

\appendix

\label{sec:appendix}

\subsection{Implementation details}
\label{appendix:implementation_details}

Table~\ref{tab:model_param} shows the employed hyperparameters for the TransCoder model. 

\begin{table}[t!]
    \centering
    \caption{The table of hyperparameters for the TransCoder modules.}
\label{tab:model_param}
\begin{tabular}{@{}llr@{}} 
 \toprule
 Notation & Hyperparameter & Value \\  \midrule
 $m$ & Block size & 3 \\ 
 $n_{heads}$ & Number of heads for self-attention & 1 \\
 $d_{khead}$ & Number of features per head & 16 \\
  $d_{model}$ & Model size $n_{heads} \times d_{khead}$ & 16 \\ 
 $\bar{N}_{enc}$ & Number of layers in the encoder & 2 \\
 $\bar{N}_{dec}$ & Number of layers in the decoder & 3 \\
 - & Total number of epochs & $10000$ \\
 - & Batches per epochs & $1$ \\
 - & Batch size & $1000$ \\
 - & Learning rate & $1e-3$ \\
 - & Scheduler & $\frac{1 - epoch}{N \ epochs}$ \\
 - & Optimizer & Adam \\
 \bottomrule
\end{tabular}
\end{table}

\subsubsection{Block Attention Mechanism} \label{sec:appendix_bam}
As mentioned previously, the TransCoder encoder $E_{T}$ and decoder $D_{T}$ (along with the refinement iteration decoder $D^{rf}_{T}$) have a symmetric structure. Similarly to the mechanism in \cite{gbaf, baaf}, the input sequence to any neural module is split into disjoint blocks of equal size. Each block is then processed and transformed into a new block of required size, while cross-block attention mechanisms capture dependencies between blocks. Each neural module comprises four main consecutive stages: pre-processing, feature extractor $FE$, positional encoding $PE$, attention feature module $AF$, sequence-to-sequence encoder $S2S$ and the final mapping layer $FMAP$. Below, we describe the structure of the neural module and the role of its components in the encoding and decoding tasks.

\textbf{Pre-Processing Stage:} The input to the neural encoder or decoder is a sequence of modulated bits (for the encoder) or noisy channel outputs (for the decoder) of length $n$, which are divided into chunks of size $m$. If the sequence length $n$ is not divisible by $m$, the input sequence is first padded with $m - n \Mod m$ ones (which correspond to zeros exposed to the BPSK-based pre-modulation) to ensure divisibility. The padded sequence is then split into $\lceil \frac{n}{m}\rceil$ blocks, each of size $m$. For the case of the refinement iteration decoder, the input consists of two vectors of size $n$: noisy channel outputs and soft estimates of probabilities per bit by the channel decoder. Then both vectors are pre-processed separately in the same way as for the single vector case, and the resulting $\lceil \frac{n}{m}\rceil$ blocks for each of the input vectors are concatenated together to form new $\lceil \frac{n}{m}\rceil$ blocks of size $2m$.     % You are correct. (c) Anastasiia

\textbf{Feature Extractor:} This part for both the encoder and decoder consists of three fully connected neural network layers, each employing a ReLU activation function. Feature extractor $FE$ transforms $m$-dimensional input blocks into blocks of dimension $d_{model}$, enabling the extraction of meaningful features. At the encoder, $FE$ captures features from the modulated codeword blocks, while at the decoder, it captures both the noise scale and the available information from the noisy input to facilitate accurate signal reconstruction. 

\textbf{Positional Encoder:}  The additional fixed positional encoding after feature extraction helps to capture the positional features that can be especially important for capturing codeword structure while using the block attention mechanism. In our model, a fixed positional embedding is used, calculated for even and odd positions correspondingly as:
\begin{equation*}
    pe_{i, 2j} = \sin{\frac{i}{1000^{\frac{2j}{L}}}}
\end{equation*}
\begin{equation*}
    pe_{i, 2j+1} = \cos{\frac{i}{1000^{\frac{2j+1}{L}}}} \ ,
\end{equation*}
where $L=200, i\in [n_{blocks}]$.

\textbf{Attention Feature Module:}  This module rescales the input of every layer in the sequence-to-sequence encoder by applying an attention mechanism between the input and the noise level similar to~\cite{xu2021wireless}. For the conventional transmission scheme, the noise standard deviation is required for LLR calculations in Eq. (\ref{eq:llrs}), then the usage of this parameter does not limit the scheme's application. First, the block of model size $d_{model}$ is concatenated with an additional channel information and passed through 2 linear layers, to obtain the mask with the same shape $d_{model}$ as an input block and then the input is scaled with weights from the obtained mask.

\textbf{Sequence-to-Sequence Encoder:} This submodule processes a sequence of $\lceil \frac{n}{m}\rceil$ number of $d_{model}$-dimensional blocks. Sequence-to-sequence encoder $S2S$ is transformer-based and consists of $N_{layers}$ identical transformer layers, where each layer consists of feed-forward network, multi-head attention, and layer normalization. Structurally, this submodule is identical for both the encoder and decoder; however, the decoder contains additional transformer layers to handle the increased complexity of the decoding task.

\textbf{Output Mapping Layer:}  The final component of the neural encoder and decoder architecture, called $FMAP$, transforms the $d_{model}$-dimensional blocks generated by the $S2S$ submodule into output symbols. 

\textbf{Transmitter Side:}
At the transmitter, the entire encoding pipeline discussed above consists of the concatenation of the channel encoder $E_C$ for the chosen channel code $\mathcal{C}(n,k)$, and the TransCoder encoder $E_T(\cdot, \bm{\theta}_{enc},m)$. which is a block attention-based encoder with trainable weights $\bm{\theta}_{enc}$.

\textbf{Receiver Side:} At the receiver, decoding is performed using a concatenation of the decoders. As previously discussed, the decoding process consists of neural decoding phase, represented by TransCoder decoders $D_{T}(\cdot; \bm{\theta}_{dec}, m)$ and $D^{rf}_{T}(\cdot; \bm{\theta}^{rf}_{dec}, m)$, with trainable weights $\bm{\theta}_{dec}$, $\bm{\theta}^{rf}_{dec}$, respectively, and channel code decoding phase, represented by a channel decoder $D_{C}$. The interaction between these two phases is complex, as the neural and channel decoders have mismatched input and output spaces. To address this, the transition between the transformer space and the channel decoder space is handled by the $T2D$ and $D2T$ modules, which we will discuss in further detail.

\textbf{Refinement Iteration:}  The main motivation behind the refinement iterations is to improve the predictions of the neural decoder while using the obtained estimations from the channel decoder. Thus, we added an additional block attention mechanism-based decoder $D^{rf}_{T}$ that processes both channel outputs and partially denoised sequence $\bm{y}_{cc}$ simultaneously. The output of the channel decoder, re-scaled by module $D2T$, can also be considered as a prior belief on the bit-wise probabilities that results in a significant improvement of final LLRs predictions.

\subsubsection{Polar Codes Training} 

To enhance the end-to-end trainability of our TransCoder, we approximated the sum-product function within the polar code's successive cancellation (SC) decoder using a pre-trained compact neural network during the training phase. This approximator, comprising three linear layers with Tanh activation functions, was selected for its superior differentiability, facilitating more effective backpropagation. For inference, the standard TransCoder architecture with the conventional sum-product operation was employed. 

\ifCLASSOPTIONcaptionsoff
  \newpage
\fi

% that's all folks

\bibliography{bibtex/bib/IEEEexample}

@misc{3gpp,
  title        = "Multiplexing and channel coding (Release 10) {3GPP} TS 21.101 v10.4.0.",
  author       = "{3GPP}",
  howpublished = "\url{https://www.3gpp.org/ftp/Specs/2023-06/Rel-10/21_series/21101-a40.zip}",
  year         = 2018
}

@misc{cc_database,
    title = "Database of Channel Codes and {ML} Simulation Results",
    author = "Helmling, M. and Scholl, S. and Gensheimer, F. and Dietz, T. and Kraft, D. and Ruzika, S. and Wehn, N.",
    howpublished = "\url{https://rptu.de/en/channel-codes}",
    year = 2019
}

@INPROCEEDINGS{polar,
  author={Arikan, Erdal},
  booktitle={2008 IEEE International Symposium on Information Theory}, 
  title={Channel polarization: A method for constructing capacity-achieving codes}, 
  year={2008},
  volume={},
  number={},
  pages={1173-1177},
  doi={10.1109/ISIT.2008.4595172}}

@ARTICLE{LDPC,
  author={Gallager, R.},
  journal={IRE Transactions on Information Theory}, 
  title={Low-density parity-check codes}, 
  year={1962},
  volume={8},
  number={1},
  pages={21-28},
  doi={10.1109/TIT.1962.1057683}}

@ARTICLE{bch_1,
  author={Hocquenghem, A.},
  journal={Chiffres (in French)}, 
  title={Codes correcteurs d'erreurs}, 
  year={1959},
  volume={2},
  pages={147-156}}

@ARTICLE{bch_2,
  author={Bose, R. C. and Ray-Chaudhuri, D. K.},
  journal={Information and Control}, 
  title={On A Class of Error Correcting Binary Group Codes}, 
  year={1960},
  volume={3},
  number={1},
  pages={68-79},
  doi={10.1016/s0019-9958(60)90287-4}}

@INPROCEEDINGS{turbo,
  author={Berrou, C. and Glavieux, A. and Thitimajshima, P.},
  booktitle={Proceedings of ICC '93 - IEEE International Conference on Communications}, 
  title={Near Shannon limit error-correcting coding and decoding: Turbo-codes. 1}, 
  year={1993},
  volume={2},
  number={},
  pages={1064-1070 vol.2},
  doi={10.1109/ICC.1993.397441}}

@inproceedings{LDPC_shannon,
  title={Near Shannon limit performance of low density parity check codes},
  author={MacKay, David JC and Neal, Radford M},
  journal={Electronics Letters},
  volume={33},
  number={6},
  pages={457--458},
  year={1997}
}

@ARTICLE{neural_polar_haim,
  author={Aharoni, Ziv and Huleihel, Bashar and Pfister, Henry D. and Permuter, Haim H.},
  journal={IEEE Transactions on Information Theory}, 
  title={Data-Driven Neural Polar Decoders for Unknown Channels With and Without Memory}, 
  year={2024},
  volume={70},
  number={12},
  pages={8495-8510},
  doi={10.1109/TIT.2024.3476681}}

@article{bounds_polyanskiy,
author = {Polyanskiy, Yury and Poor, H. Vincent and Verd\'{u}, Sergio},
title = {Channel Coding Rate in the Finite Blocklength Regime},
year = {2010},
issue_date = {May 2010},
publisher = {IEEE Press},
volume = {56},
number = {5},
issn = {0018-9448},
url = {https://doi.org/10.1109/TIT.2010.2043769},
doi = {10.1109/TIT.2010.2043769},
month = {may},
pages = {2307–2359},
numpages = {53}
}

@ARTICLE{rayleigh,
  author={Jilei Hou and Siegel, P.H. and Milstein, L.B.},
  journal={IEEE Journal on Selected Areas in Communications}, 
  title={Performance analysis and code optimization of low density parity-check codes on Rayleigh fading channels}, 
  year={2001},
  volume={19},
  number={5},
  pages={924-934},
  doi={10.1109/49.924876}}

@INPROCEEDINGS{on_deeplearning_based_cc,
  author={Gruber, Tobias and Cammerer, Sebastian and Hoydis, Jakob and Brink, Stephan ten},
  booktitle={2017 51st Annual Conference on Information Sciences and Systems (CISS)}, 
  title={On deep learning-based channel decoding}, 
  year={2017},
  volume={},
  number={},
  pages={1-6},
  doi={10.1109/CISS.2017.7926071}}

@misc{AttentionIsAllYouNeed,
  doi = {10.48550/ARXIV.1706.03762},
  url = {https://arxiv.org/abs/1706.03762},
  author = {Vaswani, Ashish and Shazeer, Noam and Parmar, Niki and Uszkoreit, Jakob and Jones, Llion and Gomez, Aidan N. and Kaiser, Lukasz and Polosukhin, Illia},
  title = {Attention Is All You Need},
  publisher = {arXiv},
  year = {2017},
  copyright = {arXiv.org perpetual, non-exclusive license}
}

@article{elias,
  title={Coding for noisy channels},
  author={Elias, Peter},
  journal={IRE Conv. Rec.},
  volume={3},
  pages={37--46},
  year={1955}
}

@ARTICLE{gbaf,
  author={Ozfatura, Emre and Shao, Yulin and Perotti, Alberto G. and Popović, Branislav M. and Gündüz, Deniz},
  journal={IEEE Journal on Selected Areas in Information Theory}, 
  title={All You Need Is Feedback: Communication With Block Attention Feedback Codes}, 
  year={2022},
  volume={3},
  number={3},
  pages={587-602}}

@inproceedings{baaf,
  title={Feedback is good, active feedback is better: Block attention active feedback codes},
  author={Ozfatura, Emre and Shao, Yulin and Ghazanfari, Amin and Perotti, Alberto and Popovic, Branislav and G{\"u}nd{\"u}z, Deniz},
  booktitle={ICC 2023-IEEE International Conference on Communications},
  pages={6652--6657},
  year={2023},
  organization={IEEE}
}

@inproceedings{ECCT,
title={Error Correction Code Transformer},
author={Yoni Choukroun and Lior Wolf},
booktitle={Advances in Neural Information Processing Systems},
editor={Alice H. Oh and Alekh Agarwal and Danielle Belgrave and Kyunghyun Cho},
year={2022}
}

@inproceedings{kurmukova2024friendly,
  title={Friendly Attacks to Improve Channel Coding Reliability},
  author={Kurmukova, Anastasiia and Gunduz, Deniz},
  booktitle={Proceedings of the AAAI Conference on Artificial Intelligence},
  volume={38},
  number={12},
  pages={13292--13300},
  year={2024}
}

@ARTICLE{bp_polar1,
  author={Arikan, Erdal},
  journal={IEEE Communications Letters}, 
  title={A performance comparison of polar codes and {R}eed-{M}uller codes}, 
  year={2008},
  volume={12},
  number={6},
  pages={447-449},
  doi={10.1109/LCOMM.2008.080017}}

@inproceedings{bp_polar2,
  title={Polar codes : A pipelined implementation},
  author={Erdal Arıkan},
  year={2010},
  url={https://api.semanticscholar.org/CorpusID:38221574}
}

@INPROCEEDINGS{scl_polar,
  author={Tal, Ido and Vardy, Alexander},
  booktitle={IEEE Int'l Symposium on Information Theory Proceedings}, 
  title={List decoding of polar codes}, 
  year={2011},
  volume={},
  number={},
  pages={1-5},
  doi={10.1109/ISIT.2011.6033904}}

@INPROCEEDINGS{neural_bp1,
  author={Nachmani, Eliya and Be'ery, Yair and Burshtein, David},
  booktitle={2016 54th Annual Allerton Conference on Communication, Control, and Computing (Allerton)}, 
  title={Learning to decode linear codes using deep learning}, 
  year={2016},
  volume={},
  number={},
  pages={341-346},
  doi={10.1109/ALLERTON.2016.7852251}}

@ARTICLE{neural_bp2,
  author={Nachmani, Eliya and Marciano, Elad and Lugosch, Loren and Gross, Warren J. and Burshtein, David and Be’ery, Yair},
  journal={IEEE Journal of Selected Topics in Signal Processing}, 
  title={Deep Learning Methods for Improved Decoding of Linear Codes}, 
  year={2018},
  volume={12},
  number={1},
  pages={119-131},
  doi={10.1109/JSTSP.2017.2788405}}

@INPROCEEDINGS{neural_bp_polar,
  author={Xu, Weihong and Wu, Zhizhen and Ueng, Yeong-Luh and You, Xiaohu and Zhang, Chuan},
  booktitle={2017 IEEE International Workshop on Signal Processing Systems (SiPS)}, 
  title={Improved polar decoder based on deep learning}, 
  year={2017},
  volume={},
  number={},
  pages={1-6},
  doi={10.1109/SiPS.2017.8109997}}

@inproceedings{neural_minsum,
author = {Lugosch, Loren and Gross, Warren J.},
title = {Neural Offset Min-Sum Decoding},
year = {2017},
publisher = {IEEE Press},
url = {https://doi.org/10.1109/ISIT.2017.8006751},
doi = {10.1109/ISIT.2017.8006751},
booktitle = {2017 IEEE International Symposium on Information Theory (ISIT)},
pages = {1361–1365},
numpages = {5},
}

@article{intro_to_ml_cc,
  title={An Introduction to Machine Learning Communications Systems},
  author={Tim O'Shea and Jakob Hoydis},
  journal={ArXiv},
  year={2017},
  volume={abs/1702.00832},
  url={https://api.semanticscholar.org/CorpusID:8010317}
}

@InProceedings{KOcodes,
  title = 	 {{KO} codes: Inventing nonlinear encoding and decoding for reliable wireless communication via deep-learning},
  author =       {Makkuva, Ashok V and Liu, Xiyang and Jamali, Mohammad Vahid and Mahdavifar, Hessam and Oh, Sewoong and Viswanath, Pramod},
  booktitle = 	 {Proceedings of the 38th International Conference on Machine Learning},
  pages = 	 {7368--7378},
  year = 	 {2021},
  editor = 	 {Meila, Marina and Zhang, Tong},
  volume = 	 {139},
  series = 	 {Proceedings of Machine Learning Research},
  month = 	 {18--24 Jul},
  publisher =    {PMLR},
  url = 	 {https://proceedings.mlr.press/v139/makkuva21a.html}
}

@inproceedings{neuralbcjr,
title={Communication Algorithms via Deep Learning},
author={Hyeji Kim and Yihan Jiang and Ranvir B. Rana and Sreeram Kannan and Sewoong Oh and Pramod Viswanath},
booktitle={International Conference on Learning Representations},
year={2018},
url={https://openreview.net/forum?id=ryazCMbR-},
}

@article{syndrom_based,
author = {Bennatan, Amir and Choukroun, Yoni and Kisilev, Pavel},
year = {2018},
month = {02},
pages = {},
title = {Deep Learning for Decoding of Linear Codes - A Syndrome-Based Approach}
}

@misc{aecct,
      title={Accelerating Error Correction Code Transformers}, 
      author={Matan Levy and Yoni Choukroun and Lior Wolf},
      year={2024},
      eprint={2410.05911},
      archivePrefix={arXiv},
      primaryClass={cs.LG},
      url={https://arxiv.org/abs/2410.05911}, 
}

@misc{crossmpt,
      title={{CrossMPT}: Cross-attention Message-Passing Transformer for Error Correcting Codes}, 
      author={Seong-Joon Park and Hee-Youl Kwak and Sang-Hyo Kim and Yongjune Kim and Jong-Seon No},
      year={2024},
      eprint={2405.01033},
      archivePrefix={arXiv},
      primaryClass={cs.LG},
      url={https://arxiv.org/abs/2405.01033}, 
}

@ARTICLE{prob_geom_shaping,
  author={Qu, Zhen and Djordjevic, Ivan B.},
  journal={IEEE Access}, 
  title={On the Probabilistic Shaping and Geometric Shaping in Optical Communication Systems}, 
  year={2019},
  volume={7},
  number={},
  pages={21454-21464},
  doi={10.1109/ACCESS.2019.2897381}}

@INPROCEEDINGS{Jones:ECOC:19,
  author={Jones, Rasmus T and Yankov, Metodi P and Zibar, Darko},
  booktitle={45th European Conference on Optical Communication (ECOC 2019)}, 
  title={End-to-end learning for GMI optimized geometric constellation shape}, 
  year={2019},
  volume={},
  number={},
  pages={1-4},
  doi={10.1049/cp.2019.0886}}

@INPROCEEDINGS{Gumus:OFC:20,
  author={Gümüş, Kadir and Alvarado, Alex and Chen, Bin and Häger, Christian and Agrell, Erik},
  booktitle={2020 Optical Fiber Communications Conference and Exhibition (OFC)}, 
  title={End-to-End Learning of Geometrical Shaping Maximizing Generalized Mutual Information}, 
  year={2020},
  volume={},
  number={},
  pages={1-3},
  doi={}}

@inproceedings{bp_standart,
    author = {{ETSI EN 302 307}},
    title = {Digital Video Broadcasting ({DVB}) standard},
    booktitle = {Second Generation Framing Structure Channel Coding and Modulation Systems for Broadcasting},
    year = {2004}
}

@INPROCEEDINGS{Aref:OFC:22,
  author={Aref, Vahid and Chagnon, Mathieu},
  booktitle={2022 Optical Fiber Communications Conference and Exhibition (OFC)}, 
  title={End-to-End Learning of Joint Geometric and Probabilistic Constellation Shaping}, 
  year={2022},
  volume={},
  number={},
  pages={1-3},
  doi={}}

@inproceedings{tinyturbo,
author = {Hebbar, S. Ashwin and Mishra, Rajesh K. and Ankireddy, Sravan Kumar and Makkuva, Ashok V. and Kim, Hyeji and Viswanath, Pramod},
title = {TinyTurbo: Efficient Turbo Decoders on Edge},
year = {2022},
publisher = {IEEE Press},
url = {https://doi.org/10.1109/ISIT50566.2022.9834589},
doi = {10.1109/ISIT50566.2022.9834589},
booktitle = {2022 IEEE International Symposium on Information Theory (ISIT)},
pages = {2797–2802},
numpages = {6},
location = {Espoo, Finland}
}

@article{Yuncheng:arXiv:24,
      title={On the Design and Performance of Machine Learning Based Error Correcting Decoders}, 
      author={Yuncheng Yuan and Péter Scheepers and Lydia Tasiou and Yunus Can Gültekin and Federico Corradi and Alex Alvarado},
      year={2024},
      journal={arXiv:eess.SP:2410.15899},
      online ={https://arxiv.org/abs/2410.15899}
}

@article{xu2021wireless,
  title={Wireless image transmission using deep source channel coding with attention modules},
  author={Xu, Jialong and Ai, Bo and Chen, Wei and Yang, Ang and Sun, Peng and Rodrigues, Miguel},
  journal={IEEE Transactions on Circuits and Systems for Video Technology},
  volume={32},
  number={4},
  pages={2315--2328},
  year={2021},
  publisher={IEEE}
}
\bibliographystyle{IEEEtran}
\end{document}